\documentclass[sigconf]{acmart}

\usepackage{subcaption}

\usepackage{natbib}
\usepackage{xspace}
\usepackage{booktabs} 
\usepackage[ruled]{algorithm2e} 
\usepackage{bm}

\SetAlFnt{\small}
\SetAlCapFnt{\small}
\SetAlCapNameFnt{\small}
\SetAlCapHSkip{0pt}
\IncMargin{-\parindent}
\usepackage{subcaption}

\newcommand{\norm}{\|}

\newcommand{\arrbegin}{\begin{eqnarray}}
\newcommand{\arrend}{\end{eqnarray}}

\newcommand{\del}{\partial}
\newcommand{\RR}{\mathbb{R}}

\usepackage{mathtools}
\usepackage[T1]{fontenc}

\DeclarePairedDelimiter\abs{\lvert}{\rvert}

\newcommand{\diag}{\textrm{diag}}

\newcommand{\bw}{\bm{w}}
\newcommand{\bv}{\bm{v}}

\newcommand{\be}{\bm{e}}

\newcommand{\bs}{\bm{s}}
\newcommand{\bz}{\bm{z}}
\newcommand{\bu}{\bm{u}}
\newcommand{\by}{\bm{y}}
\newcommand{\bx}{\bm{x}}

\newcommand{\bpi}{\bm{\pi}}

\usepackage{hyperref}  

\usepackage[capitalize,noabbrev]{cleveref}

\newtheorem{theorem}{Theorem}[section]
\newtheorem{thrm}[theorem]{Theorem}
\newtheorem{cor}[theorem]{Corollary}

\newtheorem{prop}[theorem]{Proposition}
\newtheorem{proposition}[theorem]{Proposition}
\newtheorem{defn}[theorem]{Definition}
\newtheorem{definition}[theorem]{Definition}

\newcommand{\one}{\mathbf {1}}

\newcommand{\TT}{\mathbb T}

\newcommand{\pom}{\mathsf{PoM}}

\usepackage{color-edits}

\addauthor{mp}{red}
\addauthor{aj}{blue}

\crefname{prop}{Proposition}{Propositions}

\AtBeginDocument{%
  }

\copyrightyear{2026}
\acmYear{2026}
\setcopyright{cc}
\setcctype{by}
\acmConference[WWW '26]{Proceedings of the ACM Web Conference 2026}{April 13--17, 2026}{Dubai, United Arab Emirates}
\acmBooktitle{Proceedings of the ACM Web Conference 2026 (WWW '26), April 13--17, 2026, Dubai, United Arab Emirates}
\acmPrice{}
\acmDOI{10.1145/3774904.3792080}
\acmISBN{979-8-4007-2307-0/2026/04}

\begin{document}

\title{Opinion Dynamics with Multiple Adversaries}

\author{Akhil Jalan}
\authornotemark[1]
\email{akhiljalan0@gmail.com}
\orcid{0000-0003-0148-082X}
\affiliation{%
  \institution{University of Texas at Austin}
  \department{Department of Computer Science}
  \city{Austin}
  \state{Texas}
  \country{United States}
}

\author{Marios Papachristou}
\authornote{Corresponding author.}
\email{mpapachr@asu.edu}
\orcid{0000-0002-1728-0729}
\affiliation{%
  \institution{Arizona State University}
  \department{W. P. Carey School of Business}
  \city{Tempe}
  \state{AZ}
  \country{United States}
}
\affiliation{%
  \institution{Cornell University}
  \department{Department of Computer Science}
  \city{Ithaca}
  \state{NY}
  \country{United States}
}

\authornote{Both authors contributed equally to this research}

\renewcommand{\shortauthors}{Jalan and Papachristou}

\begin{abstract}
Opinion dynamics models how the publicly expressed opinions of users in a social network coevolve according to their neighbors as well as their own intrinsic opinion. Motivated by the real-world manipulation of social networks during the 2016 US elections and the 2019 Hong Kong protests, a growing body of work models the effects of a {\em strategic actor} who interferes with the network to induce disagreement or polarization. We lift the assumption of a single strategic actor by introducing a model in which {\em any subset} of network users can manipulate network outcomes. They do so by acting according to a fictitious intrinsic opinion. Strategic actors can have {\em conflicting goals}, and push competing narratives. We characterize the Nash Equilibrium of the resulting meta-game played by the strategic actors. Experiments on real-world social network datasets from Twitter, Reddit, and Political Blogs show that strategic agents can significantly increase polarization and disagreement, as well as increase the ``cost'' of the equilibrium. To this end, we give worst-case upper bounds on the Price of Misreporting (analogous to the Price of Anarchy).  Finally, we give efficient learning algorithms for the platform to (i) detect whether strategic manipulation has occurred, and (ii) learn who the strategic actors are. Our algorithms are accurate on the same real-world datasets, suggesting how platforms can take steps to mitigate the effects of strategic behavior.

\end{abstract}



\received{20 February 2007}
\received[revised]{12 March 2009}
\received[accepted]{5 June 2009}

\maketitle

\section{Introduction}

Over the past decade, social media platforms have grown dramatically in both scale and societal influence, enabling billions of users to share information and opinions instantaneously. Online social networks now play a central role in how individuals consume news, form political views, and engage with issues related to health, products, and public life \citep{backstrom2012four,young2006diffusion,banerjee2013diffusion,shearer2021news}. 

A growing body of evidence suggests, however, that these platforms can exacerbate polarization by shaping and amplifying patterns of social interaction \citep{musco2018minimizing,chen2021adversarial,wang2024relationship,gaitonde2020adversarial}. One prominent explanation is the \emph{filter-bubble} theory \citep{pariser2011filter}, whereby personalized algorithms selectively expose users to content aligned with their past behavior and beliefs. This form of ``invisible algorithmic editing'' limits exposure to diverse viewpoints, reinforcing existing opinions and potentially undermining democratic discourse by reducing engagement with challenging or unfamiliar ideas.

Beyond endogenous polarization, social networks are increasingly subject to strategic manipulation by malicious actors seeking to induce discord. Notable examples include the alleged interference of the Russian Internet Research Agency in the 2016 U.S. presidential election \citep{mueller2018united}, coordinated attempts to disrupt protest organizations in Hong Kong \citep{twitter2019hongkong}, and large-scale scams exploiting political and conspiratorial content to maximize engagement and profit \citep{wired2024profiteers}. These interventions leverage the economics of attention on digital platforms, where inflammatory content is rewarded and algorithmically amplified, making social networks particularly vulnerable to strategic misuse.

To study the evolution of beliefs in such settings, researchers across computer science, sociology, and statistics have relied on models of \emph{opinion dynamics}, in which agents’ expressed opinions coevolve over a weighted network by combining intrinsic beliefs with social influence from neighbors \citep{Friedkin1990}. While this framework captures the interplay between individual predispositions and network effects, existing work largely focuses on scenarios involving a single adversarial or coordinating actor acting on the network to induce polarization or disagreement \citep{musco2018minimizing,chen2021adversarial,wang2024relationship,tsourakakis-2024,gaitonde2020adversarial,racz2023towards,Chitra2020}.

\begin{figure}
    \centering
    \includegraphics[height=1.5in]{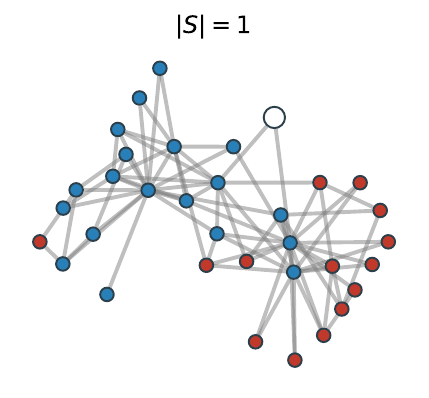}
    \includegraphics[height=1.5in]{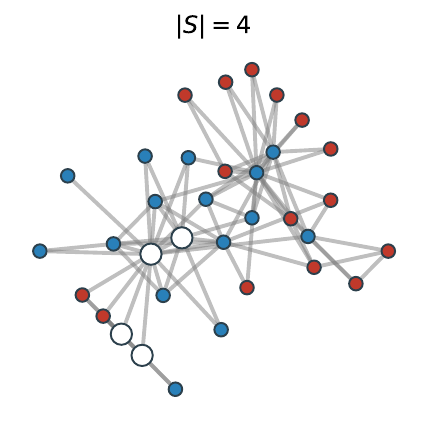}

    \caption{Visualization of the strategic equilibrium ($\bz^\prime$) on the Karate Club Graph for two different choices of $S$. The truthful intrinsic opinions have been taken to be $\bs = \bu_2$ where $\bu_2$ is the Fiedler eigenvector of $G$. The white nodes correspond to the nodes in $S$. For the other nodes, the nodes colored in blue (resp. red) correspond to nodes whose public opinion $\bz^\prime_i$ increased (resp. decreased), i.e., $(\bz^\prime_i - \bz_i) / \bz_i \ge 0$ (resp. $(\bz^\prime_i - \bz_i) / \bz_i < 0$) after $\bs'$ was chosen.}
    \label{fig:visualization}
\vspace{-2em}
\end{figure}

In this work, we lift the assumption of requiring a single actor (such as the platform) to act as an adversary to induce polarization or disagreement and consider the case of several decentralized actors. It is known that empirically, a very small percentage ($25\%$) of the users in a network need to disagree to sway consensus \citep{centola2018experimental}. Moreover, real-world social networks involve {\em multiple } malicious actors, who use different levels of manipulation and hate speech based on their individual goals~\citep{wired2024profiteers}. In this paper, we attempt to provide a theoretical basis for this phenomenon: specifically, in our setting, we assume that there is a set $S \subseteq V$ of strategic agents whose goal is to report false intrinsic opinions ($\bs'$) that are different from their true intrinsic opinions ($\bs \neq \bs'$). Their goal is to influence others while not deviating much from their neighbors; namely, they want to reach an equilibrium where their neighbors agree with them. 

For instance, assume a social network where a set of $S$ of political actors want the network to believe that their stance on a topic (e.g., abortion, elections, drug legalization, etc.) is the best. They achieve this by adversarially reporting different intrinsic opinions. This ensures that their influence is both persuasive and credible within the local network context. Such adversarial behavior can result in significantly different (cf. \cref{fig:visualization}) and highly polarized equilibria, where the strategic agents' opinions appear dominant despite not reflecting the actual intrinsic views of the majority. 

Our work investigates the conditions under which these strategic manipulations are successful, the extent of their impact on network-wide opinion dynamics, and how platforms can learn from observing these manipulated equilibria to mitigate such impacts.

In this paper, we ask the following research question (RQ): 

\begin{quote}
    \textit{\textbf{(RQ)} What if a set of strategic actors with \textbf{possibly conflicting goals} tries to manipulate the consensus by strategically reporting beliefs different than their true beliefs?}
\end{quote}

We rely on the Friedkin-Johnsen (FJ) model \citep{Friedkin1990}, where the opinions of agents coevolve via the help of a weighted undirected network $G = (V = [n], E, w)$ with non-negative weights. According to the FJ model, the agents possess intrinsic opinions $s$ and express opinions $z$, which they update via the following rule for each agent $i$:  

\begin{equation}\label{eq:fj_dynamics}
    \bz_i(t + 1) = \frac {\alpha_i \bs_i + (1 - \alpha_i) \sum_{i \sim j} w_{ij} \bz_j(t)} {1 + \sum_{i \sim j} w_{ij}}.
\end{equation}

where $\alpha_i \in (0, 1)$ is $i$'s susceptibility to persuation \citep{abebe2018opinion}. We additionally define $\Tilde \alpha_i = \alpha_i / (1 - \alpha_i)$ to be the normalized susceptibility parameter corresponding to $i$. This update rule corresponds to the best-response dynamics arising from minimizing the quadratic cost function for each $i$ \citep{bindel2011,abebe2018opinion}:

\begin{equation} \label{eq:cost_fcn}
    c_i(\bz_i, \bz_{-i}) = (1 - \alpha_i) \sum_{i \sim j} w_{ij} (\bz_i - \bz_j)^2 + \alpha_i (\bz_i - \bs_i)^2.
\end{equation}

The Pure Strategy Nash Equilibrium (PSNE) can be written as $z = ((I - A)L + A)^{-1} A \bs = B \bs$ where $L$ is the Laplacian of graph $G$, $A = \mathrm{diag}(\alpha_1, \dots, \alpha_n)$ is the diagonal matrix of susceptibilities. When an external \textit{single actor aims to induce disagreement or polarization} -- see, e.g., \citet{gaitonde2020adversarial,racz2022towards,musco2018minimizing} -- the adversary is coined with optimizing the objective function 

\begin{equation*}
    \sum_{i \in [n]} c_i(\bz_i, \bz_{-i}) = \bs^T ((I - A)L + A)^{-1} A f(L) ((I - A)L + A)^{-1} A \bs,
\end{equation*}

where $f(L)$ is a function of the Laplacian of $G$, either with optimizing towards $\bs$ \citep{gaitonde2020adversarial}, or the graph itself \citep{musco2018minimizing,racz2022towards}. 

Usually, as we also discussed earlier, many adverse actions on social networks come from \textit{several independent strategic adversaries} who try to manipulate the network by infiltrating intrinsic opinions $\bs_i^\prime$, which are different from their true stances $\bs_i$ but are simultaneously close to $\bs_i$. Unlike previous works, these ``adversaries'' can have conflicting goals. 
\renewcommand{\bs}{\bm{s}}
\renewcommand{\RR}{\mathbb{R}}

Concretely, the true opinions of the agents are $\bs_1, ..., \bs_n \in \RR$, and there is a set $S$ of deviating agents who report $\{ \bs_i^\prime \}_{i \in S}$. While an agent $i \in S$ has a true intrinsic opinion $\bs_i$, she may report some $\bs_i^\prime \neq \bs_i$ for the purpose of manipulating the dynamics of \cref{eq:fj_dynamics}. 
Specifically, her goal is to minimize her unique cost function, given by \cref{eq:cost_fcn}, at the resulting consensus $z^\prime = ((I - A)L + A)^{-1} A \bs^\prime$ where $\bs^\prime$ 
is the vector which has entries $\bs_i$ for all $i \notin S$ and $\bs_i^\prime$ for all $i \in S$. The local optimization of agent $i$ becomes:

\begin{equation} \label{eq:strategic_cost}
    \min_{\bs_i^\prime \in \RR} c_i \left (\bz^\prime = ((I - A)L + A)^{-1} A \bs^\prime \right ).
\end{equation}

Note that while \cref{eq:strategic_cost} depends on the choices of all $j \in S$, agent $i$ can only choose the $i^{th}$ entry of $\bs^\prime$. 

\noindent Our {contributions} are as follows. 

\paragraph{Characterizing Nash Equilibria with Multiple Adversaries.} We give the Nash equilibrium of the game defined by \cref{eq:strategic_cost}, and show that all Nash-optimal strategies are pure. The Pure Strategy Nash Equilibrium (PSNE) that is given by solving a constrained linear system. Given the PSNE of the game, we characterize the actors who can have the most influence in strategically manipulating the network.

\paragraph{Real-World Experiments to Understand Properites of Equilibria.} We apply our framework to real-world social network data from Twitter and Reddit \citep{Chitra2020}, and data from the Political Blogs (Polblogs) dataset \citep{adamic2005political}. We find that the influence of strategic agents can be rather significant as they can significantly increase polarization and disagreement, as well as increase the overall ``cost'' of the consensus. 

\paragraph{Analysis of Equilibrium Outcomes Under Different Sets of Strategic Actors.} Various metrics for network polarization and disagreement are sensitive to the choice of {\em who} acts strategically, in nontrivial ways. For example, adding more strategic agents can sometimes {\em decrease} the Disagreement Ratio at equilibrium (Figure~\ref{fig:ratios_number_of_deviators}), due to  counterbalancing effects. 
To address the effects of manipulation, we give worst-case upper bounds on the \textit{Price of Misreporting} (PoM), which is analogous to well-studied Price of Anarchy bounds (see, for example, \citet{bhawalkar2013,roughgarden2011local}), and suggest ways that the platform can be used to mitigate the effect of strategic behavior on their network. 

\paragraph{Learning Algorithms for the Platform.} We give  an efficient algorithm for the platform to detect if manipulation has occurred (Algorithm~\ref{alg:hypothesis_testing}), based on a hypothesis test with the publicly reported opinions $\bz^\prime$. Next, we give an algorithm to infer {\em who} manipulated the network (the set of strategic agents $S$)  from $\bz^\prime$, as long as the size of $S$ is sufficiently small. Our algorithm is inspired by the robust regression algorithm of \citet{torrent-2015}, and is practical for real-world networks. It  \textit{(i)} requires the platform to have access to node embeddings $X$ which have been shown computable even in billion-scale networks such as Twitter \citep{el2022twhin}, and \textit{(ii)} can be computed in time $(n + m)^{O(1)}$, where $n$ is the number of nodes and $m$ is the number of edges of the network. Our algorithms have high accuracy on real-world datasets from Twitter, Reddit, and Polblogs. 

\subsection{Preliminaries and Notations}

The set $[n]$ denotes $\{ 1, \dots, n \}$. $\| \bx \|_p$ denotes the $\ell_p$-norm of vector $\bx$, whose $i$-th entry is denoted by $\bx_i$. $X \succeq 0$ denotes that the matrix $X$ is positive semi-definite and $\| X \|_2$ corresponds to the spectral norm of $X$. The Laplacian of the graph $G$ is denoted by $L = D - W$ where $W$ is the weight matrix of the graph, which has entries $w_{ij} \ge 0$, and $D$ is the diagonal degree matrix with diagonal entries $D_{ii} = \sum_{i \sim j} w_{ij}$. The Laplacian has eigenvalues $0 = \lambda_1 \le \lambda_2 \le \dots \le \lambda_n$. For any undirected and connected graph $G$, $L$ is symmetric and PSD, so we can write the eigendecomposition of $L$ as: 
\begin{equation}
    L = \sum_{i \in [n]} \lambda_i \bu_i \bu_i^T \succeq 0,
\end{equation}

where $\bu_1, \dots, \bu_n$ are orthonormal eigenvectors. Moreover, $\bu_1 = (1 / \sqrt n) \one$, where $\one$ is the column vector of all 1s. $U$ denotes the matrix which has the eigenvectors of $L$ as columns; i.e., such that $L = U^T \Lambda U$ where $\Lambda = \mathrm{diag} (\lambda_1, \dots, \lambda_n)$ is the diagonal matrix of $L$'s eigenvalues. $L_i$ denotes the $i$-restricted Laplacian which corresponds to the Laplacian of the graph with all edges that are non-adjacent to $i$ being removed, and, similarly, $L_{\{ u, v \}}$ corresponds to the Laplacian of an edge $\{ u, v\}$. Note that $L_i = \sum_{i \sim j} L_{\{ i, j \}}$. For a function $f(L)$ of the Laplacian we write $f(L) = U^T f(\Lambda) U$ where $f(\Lambda) = \mathrm{diag} (f(\lambda_1), \dots, f(\lambda_n))$. For brevity, regarding the equilibrium $z$ of the FJ model, we write $B =((I - A)L + A)^{-1} A$, such that $\bz = B \bs$ and $\bz^\prime = B \bs^\prime$. Finally $\be_1, \dots, \be_n \in \RR^n$ denote the canonical basis. 

We define the total cost of an equilibrium $\bz$ to be 
\begin{equation}
    C(\bz) = \sum_{i \in [n]} c_i(\bz).
\end{equation}

We define the platform-wide metrics to be
\begin{align}
    \textrm{Polarization} \qquad \mathcal P(\bz) &= \sum_{i \in [n]} (\bz_i - \bar z)^2, \bar z = \frac 1 n \sum_{i \in [n]} \bz_i, \\
    \textrm{Disagreement} \qquad \mathcal D(\bz) &= \sum_{i, j \in [n]} w_{ij} (\bz_i - \bz_j)^2 = \bz^T L \bz. 
\end{align}

Finally, we define the \textit{``Price of Misreporting''} (PoM), which is analogous to the Price of Anarchy~\cite{roughgarden2005selfish}. The PoM is the ratio of the cost $C(\bz^\prime)$ when the agents are deviating, and the cost $C(\bz)$ when the agents are reporting truthfully, i.e.,   
\begin{equation}
    \pom := \frac {C(\bz^\prime)}{C(\bz)}.
    \label{eq:pom}
\end{equation}
Unlike the Price of Anarchy (PoA), the equilibrium $\bz$ in the denominator of Eq.~\eqref{eq:pom} is the Nash equilibrium for the Friedkin-Johnson dynamics without manipulation. In the PoA, the denominator would be $C(\bz^*)$, where $\bz^*$ is a socially optimal equilibrium~\cite{bindel2011}. 
Since we study strategic manipulations as a meta-game with respect to the base game of FJ dynamics, it is more relevant for us to compare $\bz^\prime$ with $\bz$ than with $\bz^*$. 

\subsection{Related Work}

\paragraph{Opinion Dynamics} Opinion dynamics are well-studied in computer science and economics, as well as sociology, political science, and related fields. There have been many models proposed for opinion dynamics, such as with network interactions as we study in this paper (FJ model) \citep{Friedkin1990,Bindel2015}, bounded confidence dynamics (Hegselman-Krausse Model) \citep{hegselmann2002opinion}, coevolutionary dynamics \citep{bhawalkar2013} as well as many variants of them; see, for example \citet{abebe2018opinion,hazla2019geometric,fotakis2016opinion,fotakis2023opinion,tsourakakis-2024}. The work of \citep{bindel2011} shows bounds on the Price of Anarchy (PoA) between the PSNE and the welfare-optimal solution for the FJ model, and the subsequent work of \cite{bhawalkar2013} shows PoA bounds for the coevolutionary dynamics. Additionally, the opinion dynamics have been modeled by the control community; see, for example, \citep{nedic2012,de2022,bhattacharyya2013convergence,chazelle2011total}. 

As in these works, we treat the FJ model as a basis. However, our work is significantly different as it studies a framework where any subset $S \subseteq [n]$ of strategic agents can {\em deviate} from their truthful intrinsic opinions, as opposed to studying the evolution of the expressed opinions and their PSNE in the FJ model. In our model, each strategic agent $i \in S$ can only choose a single entry $\bs_i^\prime$ of the overall deviation $\bs^\prime$, but pays a cost based on the resulting equilibrium ($\bz^\prime = B \bs^\prime$), which depends on the choices of other members of $S$. 

\paragraph{Disagreement and Polarization in Social Networks} Motivated by real-world manipulation of social networks in, e.g., the 2016 US election, a recent line of work studies polarization and strategic behavior in opinion dynamics \cite{gaitonde2020adversarial,gaitonde2021polarization,Chen2022,wang2024relationship,tsourakakis-2024,ristache2025countering}. 
\citet{Chen2022} consider a model in which an adversary can control $k \leq n$ nodes' internal opinions and seeks to maximize polarization at equilibrium. Similarly, \citet{gaitonde2020adversarial} considers a single adversary who can modify intrinsic opinions $\bm{s}$ belonging to an $\ell_2$-ball. More recent work also studies modification of agents' susceptibility parameters $\alpha_i$ to alter the median opinion at equilibrium~\cite{ristache2025countering}. Our work shares a similar perspective to these: non-strategic agents follow the base-game (in our case, Friedkin-Johnson dynamics \cref{eq:fj_dynamics}), while strategic agents treat one or more variables in the base game as a {\em strategic choice} that they may vary to manipulate game outcomes (in our case, the misreported intrinsic opinions $\{\bs_i^\prime\}_{i \in S}$). This perspective emphasizes the dynamics and equilibria of the resulting meta-game, which is played by the strategic agents against the backdrop of the base-game dynamics. Unlike these previous works, however, in our setting any subset $S \subseteq [n]$ can be strategic, and these ``adversaries'' can have conflicting goals. 

\paragraph{Manipulation of Dynamic Games.} Opinion dynamics are a widely studied instance of a network game, which is a game played by nodes in a network with payoffs depending on the actions of their neighbors~\cite{kearns2001graphical, tardos-2004}. In addition to the manipulation of opinion dynamics, researchers have studied strategic manipulation of financial network formation~\cite{jalan-chakrabarti-2024}. In the non-network setting, researchers have studied the manipulation of recommendation systmes from a game-theoretic perspective~\cite{ben2018game}, as well as security games~\cite{nguyen2019deception}, repeated auctions~\cite{kolumbus2022auctions} and Fisher markers with linear utilities~\cite{kolumbus2023asynchronous}.

\paragraph{Learning from Strategic Data.} We develop learning algorithms which observe the (possibly manipulated) equilibrium $\bz^\prime$ to detect if manipulation occurred, and if so who was responsible. The former problem relates to anomaly detection in networks. \citet{chen2022antibenford} develop a hypothesis test to detect such fraud in financial transaction neteworks, by testing if certain subgraphs deviate from Benford's Law. Similarly, \citet{agarwal2020chisel} propose a framework based on a $\chi^2$-statistic to perform graph similarity search. 

The problem of recovering the set of deviators relates to the broader literature of learning from observations of network games. Most works give learning algorithms for games {\em without} manipulation~\cite{irfan-2014,garg-2016,de2016learning,leng-2020-learning,rossi2022learning,jalan-2023}. But our data $\bz^\prime$ can be a manipulated equilibrium, which is a {\em strategic sources} of data~\cite{zampetakis2020statistics}. Learning algorithms for strategic sources are known for certain settings such as linear classifiers with small-deviation assumptions~\citep{chen2020learning}, or binary classifiers in a linear reward model~\citep{harris2023strategic}. When agents can modify their features to fool a known algorithm, even strategy-robust classifiers such as \cite{hardt2016strategic} can be inaccurate~\cite{ghalme2021strategic}.  
Since agents can deviate arbitrarily in our model, we use a robust regression method with guarantees against {\em adversarial} corruptions~\citep{torrent-2015}, similar to the learning algorithms in~\citep{kapoor2019corruption,russo2023analysis}. The work of \citet{jalan-chakrabarti-2024} studies learning from financial networks with strategic manipulations, which is in a similar spirit to our work but differs significantly in the application domain and context. 

\paragraph{Real-world Datasets}

To support our results, we use data grounded in practice, which have also been used in previous studies to study polarization and disagreement (cf. \citet{Musco2018,Chitra2020,wang2024relationship,adamic2005political}). Specifically, we use Twitter, Reddit, and Political blog networks. Both the Twitter and Reddit datasets are due to \citet{Chitra2020}. The vectors $\bs$ of initial opinions for both are obtained via sentiment analysis and also follow the post-processing of \citet{wang2024relationship}.

\paragraph{(1) Twitter dataset ($n = 548, m = 3638$).} These data correspond to debate over the Delhi legislative assembly elections of 2013. Nodes are Twitter users, and edges refer to user interactions. 

\paragraph{(2) Reddit dataset ($n = 556, m = 8969$).} These data correspond to political discussion on the \texttt{r/politics} subreddit. Nodes are who posted in the \texttt{r/politics} subreddit, and there is an edge between two users $i,j$ if two subreddits (other than \texttt{r/politics}) exist that both $i,j$ posted on during the given time period. 

\paragraph{(3) Political Blogs (Polblogs) dataset ($n = 1490, m = 16178$).} These data, due to \citet{adamic2005political}, contain opinions from political blogs (liberal and conservative). Edges between blogs were automatically extracted from a crawl of the front page of the blog. Each blog is either liberal -- where we assign a value $\bs_i = -1$ -- or conservative -- where we assign $\bs_i = +1$.

\section{Strategic Opinion Formation}\label{sec:strategic_opinion_formation}




The opinion formation game has two phases. First, strategic agents privately choose a strategic intrinsic opinion according to \cref{eq:strategic_cost}. Second, agents exchange opinions and reach consensus {\em as if} they were in the Friedkin-Johnson dynamics, except the strategic opinions are used in place of the true intrinsic opinions. 
\begin{enumerate}
    \item {\em Strategy Phase.} Each strategic agent $i \in S$ independently and privately chooses a fictitious strategic opinion $\bs_i^\prime \in \RR$. For honest agents ($i \notin S$) we have $\bs_i^\prime = \bs_i$.
    \item {\em Opinion Formation Phase.} Reach equilibrium $\bz^\prime = B \bs^\prime$ as if $\bs^\prime$ were the true intrinsic opinions~$\bs$. 
\end{enumerate}

The network $G$ and the true beliefs $\bs$ determine each agent’s utility. We pose the following problem: 

\begin{defn}[Instrinsic belief lying problem.] \label{defn:lying}
Let $S \subseteq [n]$ be a set of strategic agents. If agent $i \in S$ wants network members to express opinions close to $\bs_i$, what choice of $\bs_i^\prime$ is optimal and minimizes the cost function of \cref{eq:strategic_cost}? 
\end{defn}

The following theorem characterizes the Nash Equilibria of the Intrinsic Belief Lying Problem (cf. \cref{app:proofs} for a proof). 
\begin{theorem}[Nash Equilibrium] \label{theorem:psne}
Let $\TT_i = (1 - \alpha_i) (B^T L_i B) + \alpha_i (B^T \be_i \be_i^T B) \in \RR^{n \times n}$ and $\by_i = \alpha_i B_{ii} \bs_i$. The Nash equilibria, if any exist, are given by solutions $s^\prime \in \RR^n$ to the following constrained linear system: 
\begin{align*}
\forall i \in S: \be_i^T \TT_i \bs^\prime & = \bm{y}_i, \\
\forall j \not \in S: \bs_j^\prime & = \bs_j. 
\end{align*}
\end{theorem}

Next, we discuss some consequences of Theorem~\ref{theorem:psne}. First, we characterize $\bs^\prime$ as the solution to a linear system. 
\begin{cor} \label{cor:psne}
Let $T \in \RR^{|S| \times n}$ have rows $\{ \be_i^T \TT_i \}_{i \in S}$ given by \cref{theorem:psne}. Let $\Tilde T \in \RR^{|S| \times |S|}$ be the submatrix of $T$ selecting columns belonging to $S$. Let $\by \in \RR^{|S|}$ have entries $\by_i = \alpha_i B_{ii} \bs_i$ as above. Let $\Tilde \by = \by - \sum_{j \not \in S} \bs_j T \be_j$. Then the set of Nash equilibria, if any exist, are given by the solutions to the unconstrained linear system 

\begin{equation} \label{eq:psne}
    \Tilde T \bx = \Tilde \by.
\end{equation}
The resulting opinions vector $\bs^\prime$ is given by $\bs_i^\prime = \bx_i$ if $i \in S$ and $\bs_i^\prime = \bs_i$ otherwise.
\end{cor}

Thus, in a Nash equilibrium, every strategic agent solves their corresponding equation given by \cref{eq:psne}. The explicit characterization of equilibria also implies that Nash equilibria cannot be mixed. 
\begin{cor}[Pure Strategy Nash Equilibria]
    The Nash equilibrium corresponds to solving the system of $|S|$ linear equations in the scalars $\{ \bs_i^\prime | i \in S \}$ given by \cref{eq:psne}. Also, all Nash equilibria are pure-strategy Nash equilibria.
\end{cor}

\begin{figure}[t]
    \centering
    \includegraphics[width=0.49\linewidth]{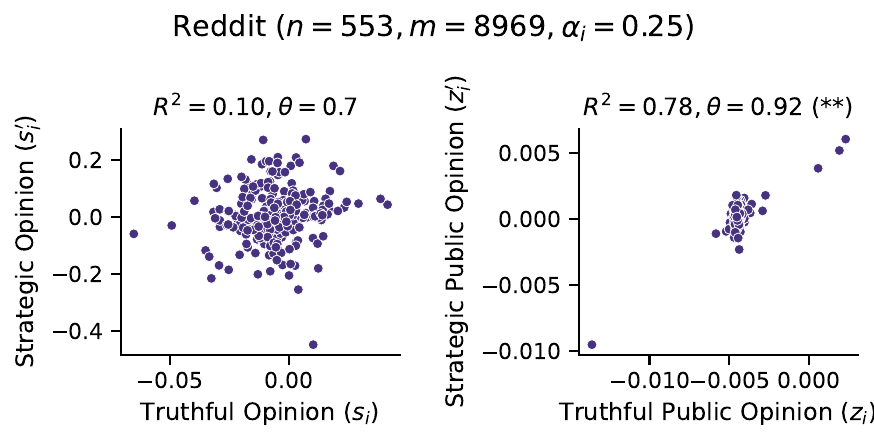} 
    \includegraphics[width=0.49\linewidth]{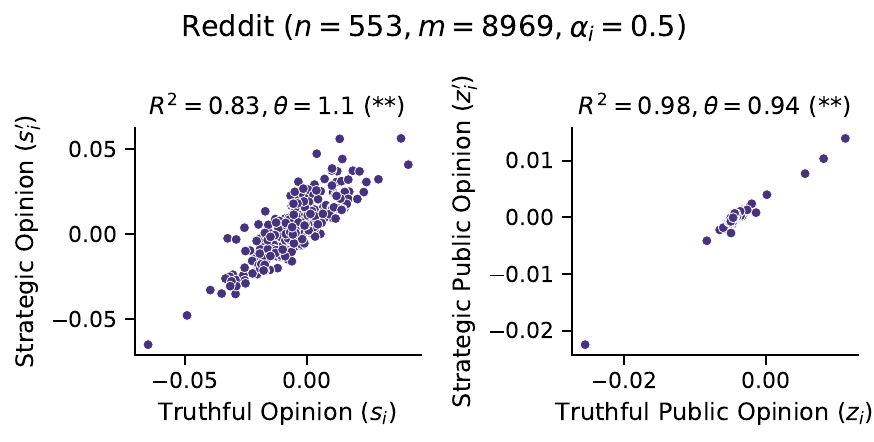} 

    \includegraphics[width=0.49\linewidth]{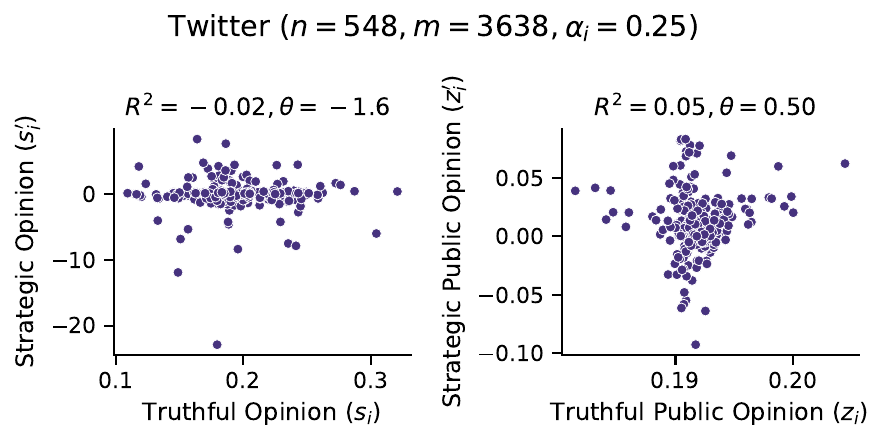}
    \includegraphics[width=0.49\linewidth]{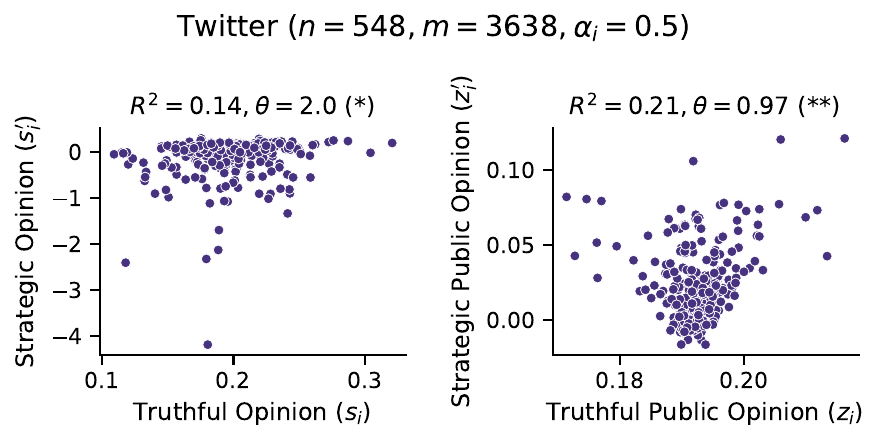}

    \includegraphics[width=0.49\linewidth]{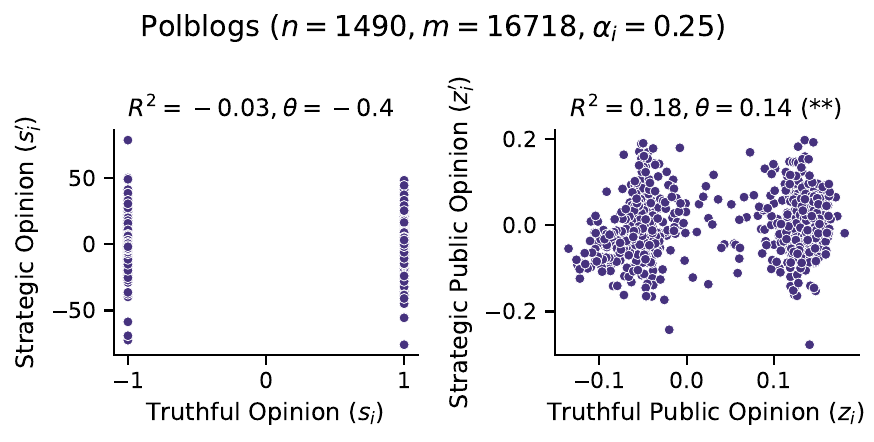} 
    \includegraphics[width=0.49\linewidth]{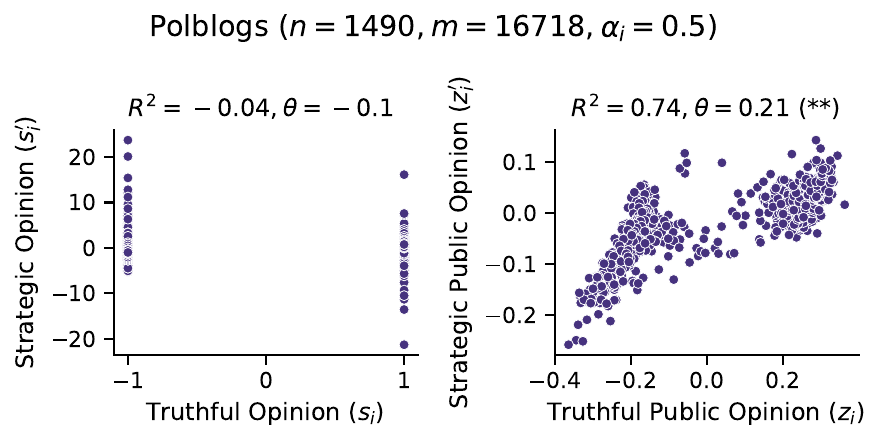} 

    \caption{Plot of truthful intrinsic opinions ($s$) and strategic opinions ($s'$), and truthful public opinions ($z$) compared to the strategic public opinions ($z'$) for the nodes belonging to $S$. $S$ is taken to be the top-50\% in terms of their eigenvector centrality. In both cases we have taken $\alpha_i \in \{ 0.25, 0.5 \}$ for all nodes. We fit a linear regression between $s'$ and $s$ (resp. between $z$ and $z'$). We report the effect size $\theta$ which corresponds to the slope of the linear regression and the $P$-value with respect to the null hypothesis ($\theta = 0$). $^{***}$ stands for $P < 0.001$, $^{**}$ stands for $P < 0.01$ and $^{*}$ stands for $P < 0.05$.}
    \label{fig:psne_data}
    \vspace{-2em}
\end{figure}

\begin{figure*}[t]
    \centering
    \includegraphics[width=0.6\linewidth]{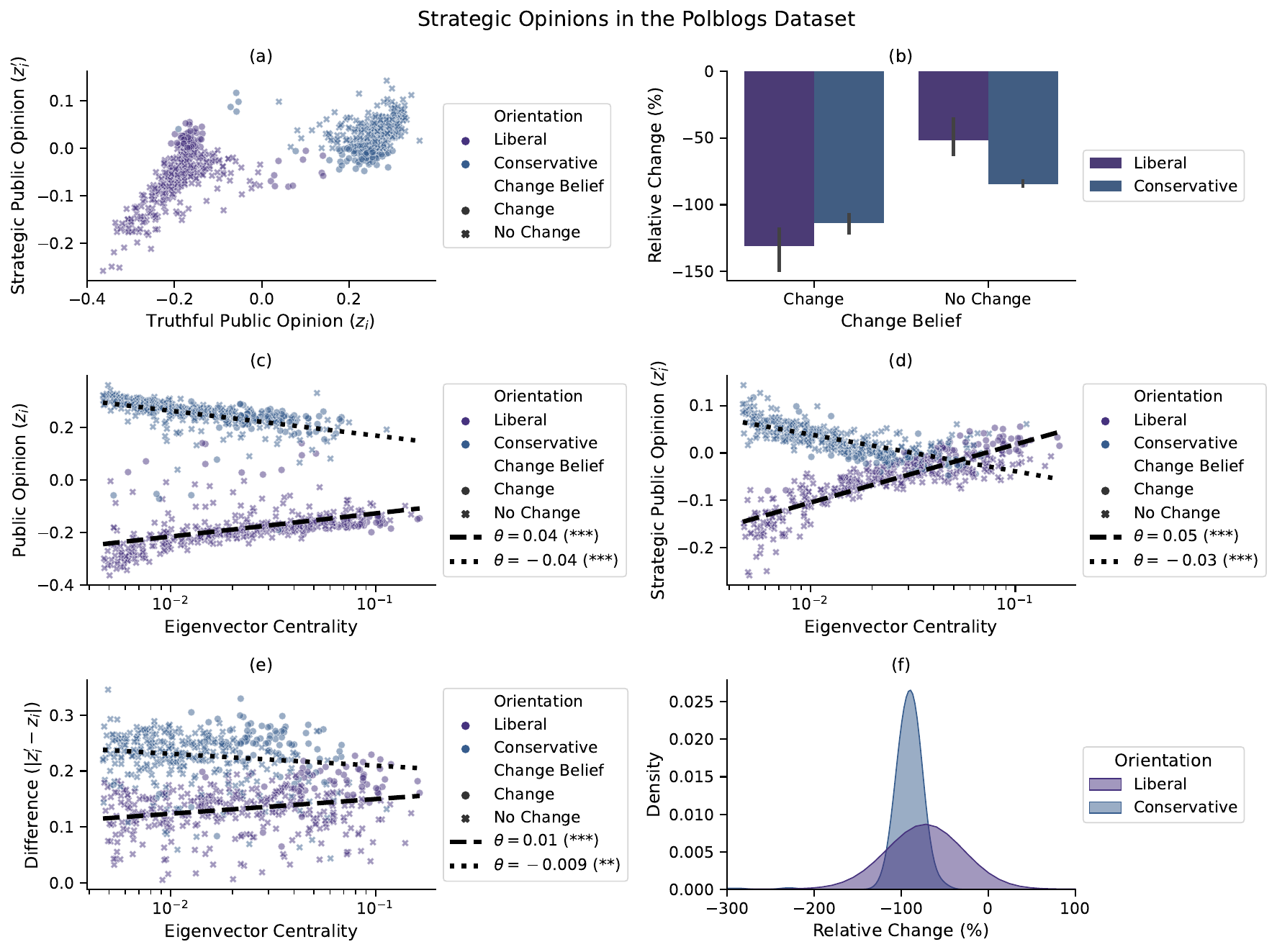}
    \caption{Strategic misreports on the Polblogs dataset, where $S$ consists of the top 50\% of agents by eigenvector centrality. Nodes are labeled as liberal ($\bs_i = -1$) or conservative ($\bs_i = +1$). A node is said to change belief if $\bz_i$ and $\bz_i'$ have opposite signs. In scatterplots (a), (c), (d), and (e), point shape indicates whether a belief change occurs, while color denotes the intrinsic opinion. We observe more frequent belief changes among liberal blogs than conservative ones (panel (b)). Panels (c) and (d) report truthful and strategic public opinions as functions of $\log \bpi_i$, while panel (e) shows the absolute change $|\bz_i' - \bz_i|$. Regression analysis reveals significant effects of $\log \bpi_i$ ($^{***}\!:\,P<0.001$, $^{**}\!:\,P<0.01$, $^{*}\!:\,P<0.05$; coefficients $\theta$) on $\bz$, $\bz'$, and $|\bz' - \bz|$, consistent with power-law structure. Finally, relative belief changes are more dispersed among liberal sources than conservative ones (panel (f)).
} 
    \label{fig:polblogs}
    \vspace{-1em}
\end{figure*}
\vspace{-1em}

\begin{figure}[t]
    \centering
    \includegraphics[width=\linewidth]{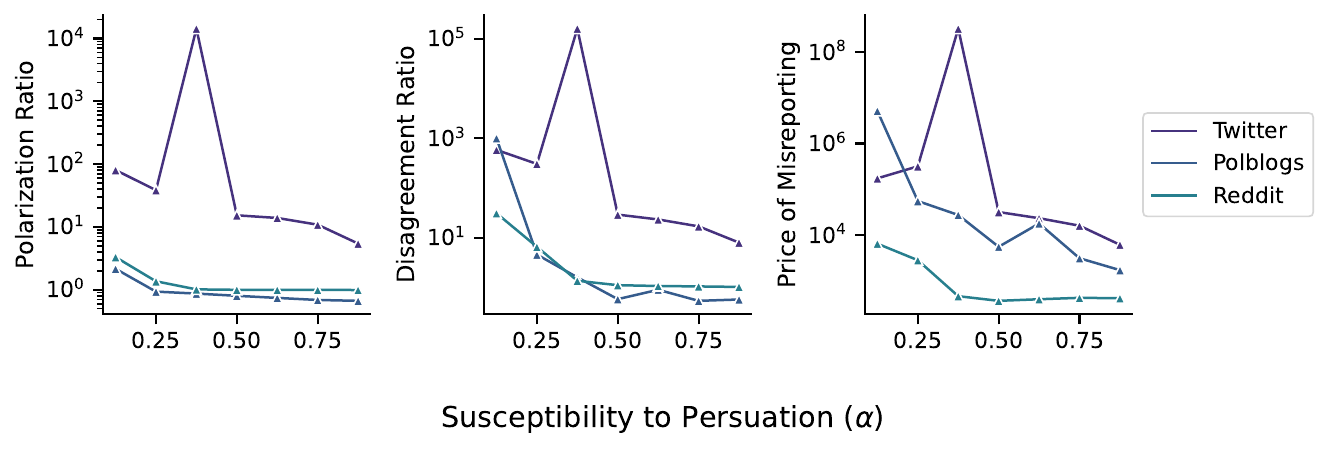}
    \caption{Polarization ratio ($\mathcal P(z')/\mathcal P(z)$), disagreement ratio ($\mathcal D(z') / \mathcal D(z)$), and price of misreporting ($C(z') / C(z)$) for the three datasets for varying susceptibility to persuasion values. We have set all susceptibilities $\alpha_i$ to the same value $\alpha$. The Twitter dataset has the largest variation in all three ratios compared to the others. $S$ is taken to be the top-50\% nodes in terms of their eigenvector centrality.}
    \label{fig:ratios_susceptibility_to_persuation}
    \vspace{-2em}
\end{figure}


\paragraph{Effect of Susceptibility to Persuasion in Real-world Data} Regarding real-world data, \cref{fig:psne_data} shows the relationship between the truthful opinions ($\bs$ and $\bz$) and the strategic ones ($\bs^\prime$ and $\bz^\prime$) for the datasets, along with the corresponding correlation coefficient $R^2$, assuming that $S$ consists of the top-50\% nodes in terms of their eigenvector centrality, for susceptibility parameters set to $\alpha_i = 0.5$ (equal self-persuasion and persuasion due to others) and $\alpha_i = 0.25$ (higher persuasion due to others). 

Regarding the public opinions, even though in the Reddit dataset, the strategic opinions seem to be correlated with the truthful ones ($R^2 = 0.78$ for $\alpha_i = 0.25$ and $R^2 = 0.94$ for $\alpha_i = 0.5$ respectively), in the Twitter dataset, we do not get the same result (i.e., $R^2 < 0.25$). Finally, in the Polblogs dataset, the situation is somewhere in the middle; when $\alpha_i = 0.25$ we get a low $R^2$ ($R^2 = 0.18$) where for $\alpha_i = 0.5$ we get a high $R^2$ ($R^2 = 0.74$). Additionally, in all cases except Twitter, we get that the effect is significant ($P < 0.01$). 

Regarding the relationship between the intrinsic opinions, we do not detect any significant effect in most cases except Reddit with $\alpha_i = 0.5$ ($P < 0.01$) and Twitter with $\alpha_i = 0.5$ ($P < 0.05$).

\paragraph{Asymmetric Effects of Strategic Behavior on Liberals and Conservatives.} \cref{fig:polblogs} analyzes the opinions of the strategic set $S$ on the Polblogs dataset. Specifically, we find that larger changes in sentiment happen across liberal outlets compared to conservative ones. Additionally, the changes in the truthful/strategic opinions are related to the eigenvector centrality $\bpi_i$ as a power law, i.e., $\bz_i' \propto \bpi_i^{\theta}$ ($P < 0.001$; linear regression between the log centralities $\log \bpi_i$ and $\bz_i'$). The same finding holds for $|\bz_i' - \bz_i|$ and $\bz_i$. 

At this point, one may wonder whether the eigenvector centrality really influences the strategic opinions $\bz_i'$ for $i \in S$. Our answer is negative. We repeat the same experiment with the Twitter and Reddit datasets, where we find no effects ($P > 0.1$; linear regression between the log centralities $\log \bpi_i$ and $\bz_i'$). Due to space limitations, the corresponding figures are deferred to \cref{app:additional_figures}. 

\paragraph{Polarization and Disagreement.} \cref{fig:ratios_susceptibility_to_persuation} shows how the polarization, disagreement, and cost change as a function of the susceptibility parameter $\alpha_i$. Except for $\alpha_i \approx 0.3$, the polarization ratio, disagreement ratio, and the price of misreporting experience a downward trend as $\alpha_i$ increases. This indicates that as as users prioritize their own opinions more than their neighbors, they are less susceptible to strategic manipulation.

\begin{figure}
    \centering
    \includegraphics[width=\linewidth]{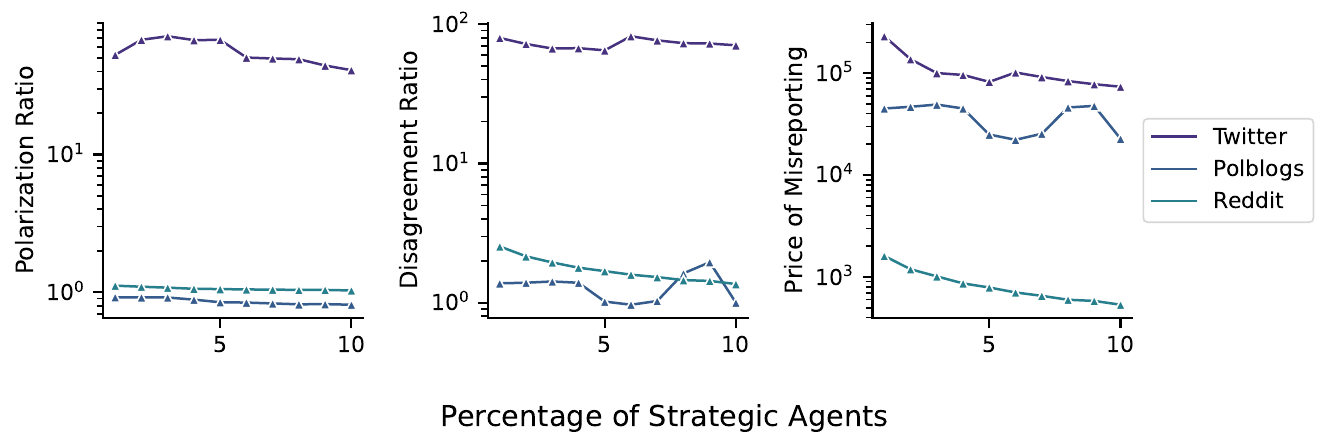}
    \caption{Polarization ratio ($\mathcal P(\bz^\prime)/\mathcal P(\bz)$), disagreement ratio ($\mathcal D(\bz^\prime) / \mathcal D(\bz)$), and price of misreporting ($C(\bz^\prime) / C(\bz)$) for the three datasets for varying the size of $|S|$. The size of $|S|$ corresponds to the top $p$ percent of the actors ($|S| = \lceil p n \rceil$) based on their eigenvector centrality (in decreasing order), for $p \in [0.01, 0.1]$. The susceptibility parameter is set to $\alpha_i = 0.5$.} 
    \label{fig:ratios_number_of_deviators}
    \vspace{-1em}
\end{figure}

\paragraph{Effect of the number of deviators ($|S|$)} Next, we study the effect of the number of deviators, which corresponds to $|S|$, on the changes in polarization, disagreement, and the total cost (through the price of misreporting). \cref{fig:ratios_number_of_deviators} shows how the polarization and disagreement when $S$ consists of the top-1-10\% most central agents with respect to eigenvector centrality. We show that even if only 1\% of agents are strategic, this can impact consensus by several orders of magnitude. 

\section{Price of Misreporting}
In Section~\ref{sec:strategic_opinion_formation}, we saw that strategic manipulation can substantially affect network outcomes via the Polarization Ratio and Disruption Ratio. We now give an upper bound for the Price of Misreporting (Eq.~\eqref{eq:pom}), which is the analogue of the Price of Anarchy in our setting. The $\pom$ measures the total cost paid by agents under the corrupted equilibrium $\bz^\prime$, versus the total cost under the non-corrupted $\bz$. Since the cost captures an agent's deviation from her {\em truthful} intrinsic opinion as well as her deviation from the expressed opinions of her neighbors, it is a natural measure of the network's discord at equilibrium. 

Theorem~\ref{thrm:pom_shared_alpha} shows that the PoM is small when the spectral radius of the Laplacian is small, and when agents are somewhat susceptible to their neighbors ($\alpha \not \to 0$). Note that the spectral radius can be replaced by a degree bound: if $d_{\textup{max}}$ is the maximum degree of the graph, then $\lambda_n \leq 2 d_{\textup{max}}$. So the PoM is small if the maximum degree is small.

\begin{theorem}
Suppose all agents deviate ($S = [n]$) and there exists $\alpha$ such that $\alpha_i = \alpha$ for all $i$.
Let $\tilde \alpha = \alpha / (1 - \alpha)$, and $\lambda_n$ be the spectral radius of the Laplacian. Then the price of misreporting is bounded as:
    \begin{align*}
        \pom \le \frac {(\lambda_n + 4 \tilde \alpha) (\lambda_n + \tilde \alpha)^2} {\tilde \alpha^5} = O \left ( \max \left \{ \frac {\lambda_n^3} {\tilde \alpha^5}, \frac {1} {\tilde \alpha^2} \right \} \right ).
    \end{align*}
\label{thrm:pom_shared_alpha}
\end{theorem}

Next, we discuss how one may generalize Theorem~\ref{thrm:pom_shared_alpha} to the case where some agents are honest: Figure~\ref{fig:ratios_susceptibility_to_persuation} shows that the PoM is {\em not} monotonic in $\abs{S}$. As the number of strategic agents grows, the PoM can fall or grow, depending on the choice of $S$, network parameters, and so on. Therefore, we would like to give a version of Theorem~\ref{thrm:pom_shared_alpha} for {\em any} set of strategic agents $S \subset [n]$, not just the case of $S = [n]$. However, proving such a bound would require analyzing $S \times S$ principal submatrices of $B, L$ to obtain characterizations of the cost at the corrupted equilibrium $\bz^\prime$. In particular, we would require a {\em restricted invertibility} estimate to prove the analogue of Eq.~\eqref{eq:ineq_Binv}. To our knowledge, the best such estimates~\citep{marcus2022interlacing} are too lossy when $n - \abs{S}$ is large. We leave this question to future work.

\section{Learning from Network Outcomes}\label{sec:learning}

To mitigate the effects of strategic behavior, a platform must understand whether manipulation has occurred, and who the strategic actors are. In this section, we give computationally efficient methods to do so based on knowledge of the network edges and observing the corrupted equilibrium $\bz^\prime$. The latter can be found, for example, by performing sentiment analysis on the users' posts. 

\paragraph{Detecting Manipulation with a Hypothesis Test}

\begin{figure}
    \centering
    \includegraphics[width=\linewidth]{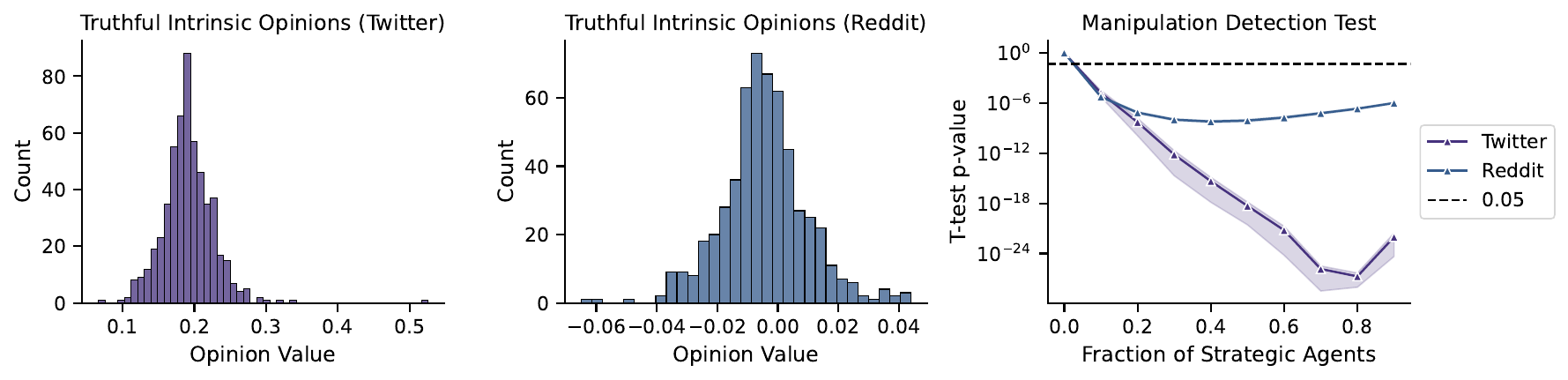}
    \caption{The true opinions for Twitter (left) and Reddit (middle) both follow a normal distribution. When simulating strategic manipulation with random choices of $S$ (right), the detection test (Algorithm~\ref{alg:hypothesis_testing}) has no Type I or Type II error at significance level $0.05$. The tester uses $\widehat{L} = L, \widehat{A} = A = \frac{1}{2} I$, and $\mu_0$ equal to the mean of the true intrinsic opinions. Shaded regions are $95\%$ confidence intervals for $p$-values of the test across $5$ independent runs.}
    \label{fig:t_test}
\end{figure}

In many real-world networks, the distribution of truthful opinions follows a Gaussian distribution (Figure~\ref{fig:t_test}). Given estimates $(\widehat{L}, \widehat{A})$ for the graph Laplacian and susceptibility matrix, the platform can observe the corrupted equilibrium $\bz^\prime$ and solve for the strategic opinions $\bs^\prime$ via: 
\begin{align}
\widehat{\bs^\prime} := \widehat{A}^{-1}((I - \widehat{A})\widehat{L} + \widehat{A}) \bz^\prime. \label{eq:solve_sprime}
\end{align}
We propose that the platform perform a one-sample $t$-test with the entries $\bs^\prime$, with a population mean $\mu_0 \in \RR$ based on e.g. historical data. Under the null hypothesis in which no manipulation has occurred, $\bs^\prime = \bs$, so the test should fail to reject the null hypothesis. However, when agents $S \subset [n]$ deviate, then $\bs_i^\prime \neq \bs_i$ for $i \in S$, so the test should reject the null for large enough deviations. The test is simple, and described in \cref{alg:hypothesis_testing}. Figure~\ref{fig:t_test} shows the results of the test for varying choices of $S$. We see that at significance level $0.05$, the test has low Type I error, as it will return ``No Manipulation'' when $S = \emptyset$, and low Type II error as it will return ``Manipulation'' when $S \neq \emptyset$.

\begin{algorithm}[t]
\SetAlgoLined
\KwIn{Estimated graph information $\widehat L$ and $\widehat A$, observed equilibrium $\bm {z}'$}
\KwOut{``Manipulation'' or ``No Manipulation''}

Observe corrupted equilibrium $\bz^\prime$. 

Solve for $\widehat{\bs^\prime}$ (Eq.~\eqref{eq:solve_sprime}).

Perform one-sample $t$ test on the entries of $\widehat{\bs^\prime}$ with a population mean $\mu_0$ under the null hypothesis. 

\Return{If the $t$-test rejects, return ``Manipulation.'' Otherwise, return ``No Manipulation.''}
\caption{Learning from Misreporting Equilibrium with Hypothesis Testing}
\label{alg:hypothesis_testing}
\end{algorithm}

For the Political Blogs dataset, intrinsic opinions belong to $\{\pm 1\}$, so the null hypothesis should be a biased Rademacher distribution. In this case, one should use a $\chi^2$-test, as in ~\citet{agarwal2020chisel,chen2022antibenford}.

\paragraph{Learning the Strategic Actors with Robust Regression} \label{sec:learning_regression}

The algorithm described in the previous section (\cref{alg:hypothesis_testing}) can be used to detect whether there exists manipulation in the network. However, the set $S$ is unknown, and therefore the platform cannot target the deviators to perform interventions to mitigate strategic behavior.  

It is, therefore, essential for the platform to be able to identify the set of deviators $S$, in case the platform needs to take regulatory actions. While at first, it may seem that finding the set of deviators $S$ is a hard task, it turns out that under mild assumptions on the intrinsic opinion formation process, we can learn the set of deviators $S$ from observing the strategically corrupted equilibrium $\bz^\prime$ via \cref{alg:learning} in polynomial time, described in \cref{alg:learning}. Our algorithm is based on robust regression leveraging the \textsc{Torrent} algorithm developed by \citet{torrent-2015} and requires access to a node embedding matrix $X \in \RR^{n \times d}$, and the size $|S|$ of the set of deviators. 

\begin{algorithm}[t]
\SetAlgoLined
\KwIn{Features $X \in \mathbb{R}^{n \times d}$, graph information $L$ and $A$, observed equilibrium $\bm {z}'$, set size $|S|$}
\KwOut{Set of strategic agents $\hat S$, estimated intrisic beliefs $\hat \bs$.}

$\widehat{\bm{s}^\prime} \leftarrow A^{-1}((I-A)L + A)\bz^\prime$\;

$\hat{v} \leftarrow $ Robust Regression (\textsc{Torrent}) with design matrix $X$, response vector $\widehat{\bm{s}^\prime}$\;

$\hat \bs \leftarrow X\hat{\bv}$\;

$\text{diffs} \leftarrow \abs{\hat \bs - \widehat{\bm{s}^\prime}}$\;

$\hat S \leftarrow \text{indices of top } k \text{ largest values in diffs}$\;

\Return{$\hat \bs \in \RR^n, \hat S \subseteq [n]$}
\caption{Learning from Misreporting Equilibrium}
\label{alg:learning}
\end{algorithm}

The key idea of \cref{alg:learning} is that if the size of the strategic set $S$ is sufficiently small, in general, $|S| \le C n$ for some small constant $C$, then we can view the misreported intrinsic opinions $\bs^\prime$ as a {\em perturbation} of the truthful opinion vector $\bs$, and then use a robust regression algorithm to estimate $\bs$. We assume that the embedding matrix $X \in \RR^{n \times d}$ determines intrinsic beliefs: for example, demography, geographic location, etc. Node-level features can be learned by a variety of methods, such as spectral embeddings on the graph Laplacian or graph neural networks. Previous works have used the framework of combining a robust estimator with model-specific information to learn from ``strategic sources'' of data, such as in bandits~\cite{kapoor2019corruption}, controls~\cite{russo2023analysis}, and network formation games~\cite{jalan-chakrabarti-2024}. 

In the sequel, we give the precise technical condition of the features required for robust regression to work \citep{torrent-2015}, which is based on conditions on the minimum and maximum eigenvalues of the correlation matrix determined by the features corresponding to agents in $S$. Specifically, for a matrix $X \in \RR^{n \times d}$ with $n$ samples in $\RR^d$ and $S \subset [n]$ let $X_S \in \RR^{\abs{S} \times d}$ select rows in $S$. Note that $\lambda_{\min}(\cdot), \lambda_{\max}(\cdot)$ are the min/max eigenvalues respectively.

\begin{definition}[SSC and SSS Conditions]
Let $\gamma \in (0,1)$. The features matrix $X \in \RR^{n \times d}$ satisfies the Subset Strong Convexity Property at level $1 - \gamma$ and Subset Strong Smoothness Property at level $\gamma$ with constants $\xi_{1-\gamma}, \Xi_\gamma$ respectively if: 
{
\begin{align*}
\xi_{1-\gamma} &\leq 
\min\limits_{S \subset [n]: \abs{S} = (1-\gamma) n} \lambda_{\min}(X_S^T X_S), \\ \Xi_{\gamma} &\geq \max\limits_{S \subset [n]: \abs{S} = \gamma n} \lambda_{\max}(X_S^T X_S).
\end{align*}}
\label{defn:ssc}
\end{definition}

\noindent We give our guarantee for \cref{alg:learning}.

\begin{proposition}
Let $X$ be as in Algorithm~\ref{alg:learning}, and suppose that $X \bv = \bs$ for some $\bv \in \RR^d$, and that $X$ satisfies the SSC condition at level $1 - \gamma$ with constant $\xi_{1-\gamma}$, and SSS condition at level $\gamma$ with constant $\Xi_\gamma$ (Definition~\ref{defn:ssc}). Then, there exist absolute constants $C, C^\prime > 0$ such that if $\abs{S} \leq Cn$ and $4\frac{\sqrt{\Xi_{\gamma}}}{\sqrt{\xi_{1-\gamma}}} < 1$, \cref{alg:learning} returns $\hat \bs$ such that $\norm \hat \bs - \bs \norm_2 \leq \norm X \norm_2 n^{-\omega(1)}$, using $T = C^\prime (\log n)^{2}$ iterations of \textsc{Torrent} for the Robust Regression step. Moreover, if for all $j \in S$ we have $\abs{\bs_j - \bs_j^\prime} \gg \norm X \norm_2 n^{-\omega(1)}$, then $\hat S = S$. 
\label{prop:b-recovery}
\end{proposition}
 
\noindent For the proof, we first state the technical result of \citet{torrent-2015} that we require. 

\begin{thrm}[\citet{torrent-2015}]
Let $X \in \RR^{n \times d}$ be a 
design matrix and $C > 0$ an absolute constant. Let $\bm b \in \{ 0, 1 \}^n$ be a corruption vector with $\norm \bm b \norm_0 \leq \alpha n$, for $\alpha \leq C$. Let $\bm{y} = X\bw^* + \bm{r}$ be the observed responses, and $\gamma \ge \alpha$ be the active set threshold given to the Algorithm 2 of \citet{torrent-2015}. Suppose $X$ satisfies the SSC property at level $1 - \gamma$ and SSS property at level $\gamma$, with constants $\xi_{1-\gamma}$ and and $\Xi_\gamma$ respectively. If the data $(X, \bm{y})$ are such that $\frac{4\sqrt{\Xi_\gamma}}{\sqrt{\xi_{1 - \gamma}}} < 1$, 
then after $t$ iterations, Algorithm 2 of \citet{torrent-2015} with active set threshold $\zeta \ge \gamma$ obtains a solution $\bw^T \in \RR^d$ such that, for large enough $n$, 
\begin{align*}
\norm \bw^T - \bw^* \norm_2 &\leq \frac{\norm \bm{r} \norm_2}{\sqrt{n}} \exp(-cT).
\end{align*}
\label{thrm:torrent}
\end{thrm}


It is interesting to investigate what an upper bound on the size of $S$ is when nodes have community memberships, such that recovery is possible, as the SSC and SSS conditions determine it. Specifically, we show the following for a blockmodel graph (proof deferred in the Appendix). 

\begin{prop} \label{prop:blockmodel}
    If $G$ has two communities with $n_1$ and $n_2$ nodes respectively such that $n_1 \ge n_2 \ge 1$, and $X \in \{ 0, 1 \}^{n \times 2}$ is an embedding vector where each row $\bx_i$ corresponds to a one-hot vector for the community of node $i$, then \cref{alg:learning} can recover $S$ perfectly as long as $|S| < \frac {n_1} {17}$.
\end{prop}

For instance, when \cref{prop:blockmodel} is applied to the Polblogs dataset, it shows that $S$ can be fully recovered as long as $|S| \le 9$. We can obtain a slightly worse bound and extend the result to a blockmodel graph with $K$ communities (proof in the Appendix). 

\begin{prop} \label{prop:blockmodel_k}
    If $G$ has $K \ge 2$ communities with sizes $n_1 \ge n_2 \ge \dots \ge n_K$  with $\left ( \frac {16K} {16K + 1} \right ) \frac n K < n_K \le \frac n K$ and $X \in \{ 0, 1 \}^{n \times K}$ is an embedding vector where $\bx_i$ corresponds to an one-hot encoding of the community membership, then \cref{alg:learning} can recover $S$ perfectly as long as $|S| < \frac {1} {16K + 1} n$.  
\end{prop}

\begin{figure}
    \centering
    \includegraphics[width=\linewidth]{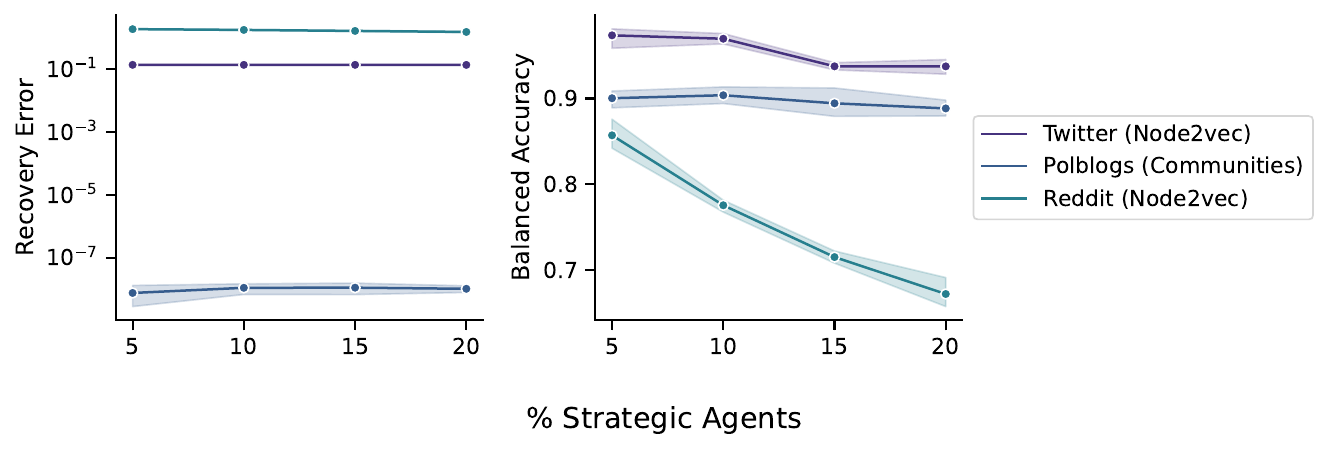}
    \caption{Reconstruction error and balanced accuracy for the robust regression problem in \cref{alg:learning} as a function of the fraction of strategic agents. The left panel reports the recovery error $\frac{1}{n}\sum_{i\in[n]}\left|\frac{\hat\bs_i-\bs_i}{\bs_i}\right|$, while the right panel shows the balanced accuracy between the recovered set $\hat S$ and the true set $S$. Confidence intervals are obtained by sampling $S$ uniformly at random from $[n]$ five times for each $|S|$. For the Twitter and Reddit datasets, we use 128-dimensional Node2vec embeddings. For Polblogs, node features correspond to community membership (liberal $\bx_i=(1,0)$, conservative $\bx_i=(0,1)$), with intrinsic opinions given by $\bs=X\bv$ for $\bv=(1,-1)^\top$. We also report results using 128-dimensional spectral embeddings. For \textsc{Torrent}, the recovery threshold is set to $|S|/n$ and the step size to $\eta = 1/\|\bar X\|_2^2$, where $\bar X$ denotes the min--max normalized embedding matrix.}
    \label{fig:learn_S}
    \vspace{-1.5em}
\end{figure}

In detail, \cref{prop:blockmodel_k} states that as long as the smallest community of the graph has size $\Theta (n/K)$ then the recovery of a set of size $|S| = O(n/K)$ is possible. If $\abs{S} \gg n/K$ and $S$ contains all members of the smallest community, then robust regression can fail.  We show that in real-world datasets, \cref{alg:learning} can identify the set of deviators with high accuracy (cf. \cref{fig:learn_S}). Specifically, in the real-world datasets we take $S$ to be a randomly sampled set of size $\lceil pn \rceil$ for $p \in \{ 0.05, 0.1, 0.15, 0.2 \}$ and the embeddings to be 128-dimenaional Node2Vec embeddings for Twitter and Reddit and community membership embeddings for the Pollblogs dataset. \cref{alg:learning} achieves low recovery error as well as high balanced accuracy score.

\section{Managerial Insights and Discussion}

\subsection{Platform Governance}

Our characterization of the Nash Equilibrium in multi-adversary opinion dynamics offers three strategic pillars for social media platform governance:

    \noindent  \textbf{PoM as a Vulnerability Metric.} The \textit{Price of Misreporting (PoM)} quantifies the social cohesion lost to strategic dishonesty. Platform stakeholders should prioritize monitoring networks with high connectivity but low individual susceptibility, as these ``viral'' environments are most vulnerable to disproportionate increases in polarization from even small-scale coordinated misreporting.
    
    \noindent \textbf{Beyond Content Takedowns.} Because strategic actors optimize reported opinions based on network structure, they can exert significant influence without triggering traditional platform health interventions (for example, flagging a post as fake news). Platforms must shift from individual post moderation to monitoring \textit{structural shifts} and community sensitivity using ``what-if'' simulations of the Nash Equilibrium.

    \noindent \textbf{Structural Friction.} To narrow the gap between reported and intrinsic opinions, platforms can effectively increase the sincerity cost by implementing friction mechanisms. Increasing transparency of historical stances and cross-network consistency makes maintaining strategic ``lies'' more costly, disincentivizing extreme misreporting.

\subsection{Conclusion}

We introduced a game-theoretic framework that replaces the "single bad actor" assumption with a realistic model of multiple, competing adversaries. We proved the existence of a unique Nash Equilibrium and introduced the Price of Misreporting to bound social inefficiency. Our empirical results across Reddit, Twitter, and Political Blogs datasets demonstrate that multi-agent competition often heightens global disagreement rather than neutralizing it. These findings provide a mathematical foundation for a new generation of intervention tools that prioritize the structural integrity of social graphs over keyword-based moderation. Future research should investigate how these strategic behaviors evolve alongside adaptive network topologies.

\begin{acks}
    The authors would like to thank Yanbang Wang, Yixuan He, Jon Kleinberg, Giannis Fikioris, and Olivia Sheng for their insightful comments on the current manuscript. 
\end{acks}

\bibliographystyle{ACM-Reference-Format}
\bibliography{main}


\begin{thebibliography}{62}


\ifx \showCODEN    \undefined \def \showCODEN     #1{\unskip}     \fi
\ifx \showISBNx    \undefined \def \showISBNx     #1{\unskip}     \fi
\ifx \showISBNxiii \undefined \def \showISBNxiii  #1{\unskip}     \fi
\ifx \showISSN     \undefined \def \showISSN      #1{\unskip}     \fi
\ifx \showLCCN     \undefined \def \showLCCN      #1{\unskip}     \fi
\ifx \shownote     \undefined \def \shownote      #1{#1}          \fi
\ifx \showarticletitle \undefined \def \showarticletitle #1{#1}   \fi
\ifx \showURL      \undefined \def \showURL       {\relax}        \fi
\providecommand\bibfield[2]{#2}
\providecommand\bibinfo[2]{#2}
\providecommand\natexlab[1]{#1}
\providecommand\showeprint[2][]{arXiv:#2}

\bibitem[Abebe et~al\mbox{.}(2018)]%
        {abebe2018opinion}
\bibfield{author}{\bibinfo{person}{Rediet Abebe}, \bibinfo{person}{Jon Kleinberg}, \bibinfo{person}{David Parkes}, {and} \bibinfo{person}{Charalampos~E Tsourakakis}.} \bibinfo{year}{2018}\natexlab{}.
\newblock \showarticletitle{Opinion dynamics with varying susceptibility to persuasion}. In \bibinfo{booktitle}{\emph{Proceedings of the 24th ACM SIGKDD International Conference on Knowledge Discovery \& Data Mining}}. \bibinfo{pages}{1089--1098}.
\newblock


\bibitem[Adamic and Glance(2005)]%
        {adamic2005political}
\bibfield{author}{\bibinfo{person}{Lada~A Adamic} {and} \bibinfo{person}{Natalie Glance}.} \bibinfo{year}{2005}\natexlab{}.
\newblock \showarticletitle{The political blogosphere and the 2004 US election: divided they blog}. In \bibinfo{booktitle}{\emph{Proceedings of the 3rd international workshop on Link discovery}}. \bibinfo{pages}{36--43}.
\newblock


\bibitem[Agarwal et~al\mbox{.}(2020)]%
        {agarwal2020chisel}
\bibfield{author}{\bibinfo{person}{Shubhangi Agarwal}, \bibinfo{person}{Sourav Dutta}, {and} \bibinfo{person}{Arnab Bhattacharya}.} \bibinfo{year}{2020}\natexlab{}.
\newblock \showarticletitle{Chisel: Graph similarity search using chi-squared statistics in large probabilistic graphs}.
\newblock \bibinfo{journal}{\emph{Proceedings of the VLDB Endowment}} \bibinfo{volume}{13}, \bibinfo{number}{10} (\bibinfo{year}{2020}), \bibinfo{pages}{1654--1668}.
\newblock


\bibitem[Backstrom et~al\mbox{.}(2012)]%
        {backstrom2012four}
\bibfield{author}{\bibinfo{person}{Lars Backstrom}, \bibinfo{person}{Paolo Boldi}, \bibinfo{person}{Marco Rosa}, \bibinfo{person}{Johan Ugander}, {and} \bibinfo{person}{Sebastiano Vigna}.} \bibinfo{year}{2012}\natexlab{}.
\newblock \showarticletitle{Four degrees of separation}. In \bibinfo{booktitle}{\emph{Proceedings of the 4th annual ACM Web science conference}}. \bibinfo{pages}{33--42}.
\newblock


\bibitem[Banerjee et~al\mbox{.}(2013)]%
        {banerjee2013diffusion}
\bibfield{author}{\bibinfo{person}{Abhijit Banerjee}, \bibinfo{person}{Arun~G Chandrasekhar}, \bibinfo{person}{Esther Duflo}, {and} \bibinfo{person}{Matthew~O Jackson}.} \bibinfo{year}{2013}\natexlab{}.
\newblock \showarticletitle{The diffusion of microfinance}.
\newblock \bibinfo{journal}{\emph{Science}} \bibinfo{volume}{341}, \bibinfo{number}{6144} (\bibinfo{year}{2013}), \bibinfo{pages}{1236498}.
\newblock


\bibitem[Ben-Porat and Tennenholtz(2018)]%
        {ben2018game}
\bibfield{author}{\bibinfo{person}{Omer Ben-Porat} {and} \bibinfo{person}{Moshe Tennenholtz}.} \bibinfo{year}{2018}\natexlab{}.
\newblock \showarticletitle{A game-theoretic approach to recommendation systems with strategic content providers}.
\newblock \bibinfo{journal}{\emph{Advances in Neural Information Processing Systems}}  \bibinfo{volume}{31} (\bibinfo{year}{2018}).
\newblock


\bibitem[Bhatia et~al\mbox{.}(2015)]%
        {torrent-2015}
\bibfield{author}{\bibinfo{person}{Kush Bhatia}, \bibinfo{person}{Prateek Jain}, {and} \bibinfo{person}{Purushottam Kar}.} \bibinfo{year}{2015}\natexlab{}.
\newblock \showarticletitle{Robust regression via hard thresholding}.
\newblock \bibinfo{journal}{\emph{Advances in neural information processing systems}}  \bibinfo{volume}{28} (\bibinfo{year}{2015}).
\newblock


\bibitem[Bhattacharyya et~al\mbox{.}(2013)]%
        {bhattacharyya2013convergence}
\bibfield{author}{\bibinfo{person}{Arnab Bhattacharyya}, \bibinfo{person}{Mark Braverman}, \bibinfo{person}{Bernard Chazelle}, {and} \bibinfo{person}{Huy~L Nguyen}.} \bibinfo{year}{2013}\natexlab{}.
\newblock \showarticletitle{On the convergence of the Hegselmann-Krause system}. In \bibinfo{booktitle}{\emph{Proceedings of the 4th conference on Innovations in Theoretical Computer Science}}. \bibinfo{pages}{61--66}.
\newblock


\bibitem[Bhawalkar et~al\mbox{.}(2013)]%
        {bhawalkar2013}
\bibfield{author}{\bibinfo{person}{Kshipra Bhawalkar}, \bibinfo{person}{Sreenivas Gollapudi}, {and} \bibinfo{person}{Kamesh Munagala}.} \bibinfo{year}{2013}\natexlab{}.
\newblock \showarticletitle{Coevolutionary opinion formation games}. In \bibinfo{booktitle}{\emph{Proceedings of the forty-fifth annual ACM symposium on Theory of computing}}. \bibinfo{pages}{41--50}.
\newblock


\bibitem[Bindel et~al\mbox{.}(2011)]%
        {bindel2011}
\bibfield{author}{\bibinfo{person}{David Bindel}, \bibinfo{person}{Jon Kleinberg}, {and} \bibinfo{person}{Sigal Oren}.} \bibinfo{year}{2011}\natexlab{}.
\newblock \showarticletitle{How Bad is Forming Your Own Opinion?}. In \bibinfo{booktitle}{\emph{2011 IEEE 52nd Annual Symposium on Foundations of Computer Science}}. IEEE, \bibinfo{pages}{57--66}.
\newblock


\bibitem[Bindel et~al\mbox{.}(2015)]%
        {Bindel2015}
\bibfield{author}{\bibinfo{person}{David Bindel}, \bibinfo{person}{Jon Kleinberg}, {and} \bibinfo{person}{Sigal Oren}.} \bibinfo{year}{2015}\natexlab{}.
\newblock \showarticletitle{How bad is forming your own opinion?}
\newblock \bibinfo{journal}{\emph{Games and Economic Behavior}}  \bibinfo{volume}{92} (\bibinfo{year}{2015}), \bibinfo{pages}{248--265}.
\newblock


\bibitem[Bjork-James and Donovan(2024)]%
        {wired2024profiteers}
\bibfield{author}{\bibinfo{person}{Sophie Bjork-James} {and} \bibinfo{person}{Joan Donovan}.} \bibinfo{year}{2024}\natexlab{}.
\newblock \bibinfo{title}{Profiteers Are Exploiting US Election Conspiracies and Hate to Make Millions}.
\newblock
\urldef\tempurl%
\url{https://www.wired.com/story/2024-election-profiteers?utm_source=chatgpt.com}
\showURL{%
\tempurl}
\newblock
\shownote{Accessed: 2024-12-27}.


\bibitem[Centola et~al\mbox{.}(2018)]%
        {centola2018experimental}
\bibfield{author}{\bibinfo{person}{Damon Centola}, \bibinfo{person}{Joshua Becker}, \bibinfo{person}{Devon Brackbill}, {and} \bibinfo{person}{Andrea Baronchelli}.} \bibinfo{year}{2018}\natexlab{}.
\newblock \showarticletitle{Experimental evidence for tipping points in social convention}.
\newblock \bibinfo{journal}{\emph{Science}} \bibinfo{volume}{360}, \bibinfo{number}{6393} (\bibinfo{year}{2018}), \bibinfo{pages}{1116--1119}.
\newblock


\bibitem[Chazelle(2011)]%
        {chazelle2011total}
\bibfield{author}{\bibinfo{person}{Bernard Chazelle}.} \bibinfo{year}{2011}\natexlab{}.
\newblock \showarticletitle{The total s-energy of a multiagent system}.
\newblock \bibinfo{journal}{\emph{SIAM Journal on Control and Optimization}} \bibinfo{volume}{49}, \bibinfo{number}{4} (\bibinfo{year}{2011}), \bibinfo{pages}{1680--1706}.
\newblock


\bibitem[Chen and R{\'a}cz(2021a)]%
        {chen2021adversarial}
\bibfield{author}{\bibinfo{person}{Mayee~F Chen} {and} \bibinfo{person}{Mikl{\'o}s~Z R{\'a}cz}.} \bibinfo{year}{2021}\natexlab{a}.
\newblock \showarticletitle{An adversarial model of network disruption: Maximizing disagreement and polarization in social networks}.
\newblock \bibinfo{journal}{\emph{IEEE Transactions on Network Science and Engineering}} \bibinfo{volume}{9}, \bibinfo{number}{2} (\bibinfo{year}{2021}), \bibinfo{pages}{728--739}.
\newblock


\bibitem[Chen and R{\'a}cz(2021b)]%
        {Chen2022}
\bibfield{author}{\bibinfo{person}{Mayee~F Chen} {and} \bibinfo{person}{Mikl{\'o}s~Z R{\'a}cz}.} \bibinfo{year}{2021}\natexlab{b}.
\newblock \showarticletitle{An adversarial model of network disruption: Maximizing disagreement and polarization in social networks}.
\newblock \bibinfo{journal}{\emph{IEEE Transactions on Network Science and Engineering}} \bibinfo{volume}{9}, \bibinfo{number}{2} (\bibinfo{year}{2021}), \bibinfo{pages}{728--739}.
\newblock


\bibitem[Chen and Tsourakakis(2022)]%
        {chen2022antibenford}
\bibfield{author}{\bibinfo{person}{Tianyi Chen} {and} \bibinfo{person}{Charalampos Tsourakakis}.} \bibinfo{year}{2022}\natexlab{}.
\newblock \showarticletitle{Antibenford subgraphs: Unsupervised anomaly detection in financial networks}. In \bibinfo{booktitle}{\emph{Proceedings of the 28th ACM SIGKDD Conference on Knowledge Discovery and Data Mining}}. \bibinfo{pages}{2762--2770}.
\newblock


\bibitem[Chen et~al\mbox{.}(2020)]%
        {chen2020learning}
\bibfield{author}{\bibinfo{person}{Yiling Chen}, \bibinfo{person}{Yang Liu}, {and} \bibinfo{person}{Chara Podimata}.} \bibinfo{year}{2020}\natexlab{}.
\newblock \showarticletitle{Learning strategy-aware linear classifiers}.
\newblock \bibinfo{journal}{\emph{Advances in Neural Information Processing Systems}}  \bibinfo{volume}{33} (\bibinfo{year}{2020}), \bibinfo{pages}{15265--15276}.
\newblock


\bibitem[Chitra and Musco(2020)]%
        {Chitra2020}
\bibfield{author}{\bibinfo{person}{Uthsav Chitra} {and} \bibinfo{person}{Christopher Musco}.} \bibinfo{year}{2020}\natexlab{}.
\newblock \showarticletitle{Analyzing the impact of filter bubbles on social network polarization}. In \bibinfo{booktitle}{\emph{Proceedings of the 13th International Conference on Web Search and Data Mining}}. \bibinfo{pages}{115--123}.
\newblock


\bibitem[De et~al\mbox{.}(2016)]%
        {de2016learning}
\bibfield{author}{\bibinfo{person}{Abir De}, \bibinfo{person}{Isabel Valera}, \bibinfo{person}{Niloy Ganguly}, \bibinfo{person}{Sourangshu Bhattacharya}, {and} \bibinfo{person}{Manuel Gomez~Rodriguez}.} \bibinfo{year}{2016}\natexlab{}.
\newblock \showarticletitle{Learning and forecasting opinion dynamics in social networks}.
\newblock \bibinfo{journal}{\emph{Advances in neural information processing systems}}  \bibinfo{volume}{29} (\bibinfo{year}{2016}).
\newblock


\bibitem[De~Pasquale and Valcher(2022)]%
        {de2022}
\bibfield{author}{\bibinfo{person}{Giulia De~Pasquale} {and} \bibinfo{person}{Maria~Elena Valcher}.} \bibinfo{year}{2022}\natexlab{}.
\newblock \showarticletitle{Multi-dimensional extensions of the Hegselmann-Krause model}. In \bibinfo{booktitle}{\emph{2022 IEEE 61st Conference on Decision and Control (CDC)}}. IEEE, \bibinfo{pages}{3525--3530}.
\newblock


\bibitem[El-Kishky et~al\mbox{.}(2022)]%
        {el2022twhin}
\bibfield{author}{\bibinfo{person}{Ahmed El-Kishky}, \bibinfo{person}{Thomas Markovich}, \bibinfo{person}{Serim Park}, \bibinfo{person}{Chetan Verma}, \bibinfo{person}{Baekjin Kim}, \bibinfo{person}{Ramy Eskander}, \bibinfo{person}{Yury Malkov}, \bibinfo{person}{Frank Portman}, \bibinfo{person}{Sof{\'\i}a Samaniego}, \bibinfo{person}{Ying Xiao}, {et~al\mbox{.}}} \bibinfo{year}{2022}\natexlab{}.
\newblock \showarticletitle{Twhin: Embedding the twitter heterogeneous information network for personalized recommendation}. In \bibinfo{booktitle}{\emph{Proceedings of the 28th ACM SIGKDD conference on knowledge discovery and data mining}}. \bibinfo{pages}{2842--2850}.
\newblock


\bibitem[Fotakis et~al\mbox{.}(2023)]%
        {fotakis2023opinion}
\bibfield{author}{\bibinfo{person}{Dimitris Fotakis}, \bibinfo{person}{Vardis Kandiros}, \bibinfo{person}{Vasilis Kontonis}, {and} \bibinfo{person}{Stratis Skoulakis}.} \bibinfo{year}{2023}\natexlab{}.
\newblock \showarticletitle{Opinion dynamics with limited information}.
\newblock \bibinfo{journal}{\emph{Algorithmica}} \bibinfo{volume}{85}, \bibinfo{number}{12} (\bibinfo{year}{2023}), \bibinfo{pages}{3855--3888}.
\newblock


\bibitem[Fotakis et~al\mbox{.}(2016)]%
        {fotakis2016opinion}
\bibfield{author}{\bibinfo{person}{Dimitris Fotakis}, \bibinfo{person}{Dimitris Palyvos-Giannas}, {and} \bibinfo{person}{Stratis Skoulakis}.} \bibinfo{year}{2016}\natexlab{}.
\newblock \showarticletitle{Opinion Dynamics with Local Interactions.}. In \bibinfo{booktitle}{\emph{IJCAI}}. \bibinfo{pages}{279--285}.
\newblock


\bibitem[Friedkin and Johnsen(1990)]%
        {Friedkin1990}
\bibfield{author}{\bibinfo{person}{Noah~E Friedkin} {and} \bibinfo{person}{Eugene~C Johnsen}.} \bibinfo{year}{1990}\natexlab{}.
\newblock \showarticletitle{Social influence and opinions}.
\newblock \bibinfo{journal}{\emph{Journal of Mathematical Sociology}} \bibinfo{volume}{15}, \bibinfo{number}{3-4} (\bibinfo{year}{1990}), \bibinfo{pages}{193--206}.
\newblock


\bibitem[Gaitonde et~al\mbox{.}(2020)]%
        {gaitonde2020adversarial}
\bibfield{author}{\bibinfo{person}{Jason Gaitonde}, \bibinfo{person}{Jon Kleinberg}, {and} \bibinfo{person}{Eva Tardos}.} \bibinfo{year}{2020}\natexlab{}.
\newblock \showarticletitle{Adversarial perturbations of opinion dynamics in networks}. In \bibinfo{booktitle}{\emph{Proceedings of the 21st ACM Conference on Economics and Computation}}. \bibinfo{pages}{471--472}.
\newblock


\bibitem[Gaitonde et~al\mbox{.}(2021)]%
        {gaitonde2021polarization}
\bibfield{author}{\bibinfo{person}{Jason Gaitonde}, \bibinfo{person}{Jon Kleinberg}, {and} \bibinfo{person}{{\'E}va Tardos}.} \bibinfo{year}{2021}\natexlab{}.
\newblock \showarticletitle{Polarization in geometric opinion dynamics}. In \bibinfo{booktitle}{\emph{Proceedings of the 22nd ACM Conference on Economics and Computation}}. \bibinfo{pages}{499--519}.
\newblock


\bibitem[Garg and Jaakkola(2016)]%
        {garg-2016}
\bibfield{author}{\bibinfo{person}{Vikas Garg} {and} \bibinfo{person}{Tommi Jaakkola}.} \bibinfo{year}{2016}\natexlab{}.
\newblock \showarticletitle{Learning Tree Structured Potential Games}. In \bibinfo{booktitle}{\emph{Advances in Neural Information Processing Systems}}, \bibfield{editor}{\bibinfo{person}{D.~Lee}, \bibinfo{person}{M.~Sugiyama}, \bibinfo{person}{U.~Luxburg}, \bibinfo{person}{I.~Guyon}, {and} \bibinfo{person}{R.~Garnett}} (Eds.), Vol.~\bibinfo{volume}{29}. \bibinfo{publisher}{Curran Associates, Inc.}
\newblock
\urldef\tempurl%
\url{https://proceedings.neurips.cc/paper_files/paper/2016/file/22ac3c5a5bf0b520d281c122d1490650-Paper.pdf}
\showURL{%
\tempurl}


\bibitem[Ghalme et~al\mbox{.}(2021)]%
        {ghalme2021strategic}
\bibfield{author}{\bibinfo{person}{Ganesh Ghalme}, \bibinfo{person}{Vineet Nair}, \bibinfo{person}{Itay Eilat}, \bibinfo{person}{Inbal Talgam-Cohen}, {and} \bibinfo{person}{Nir Rosenfeld}.} \bibinfo{year}{2021}\natexlab{}.
\newblock \showarticletitle{Strategic classification in the dark}. In \bibinfo{booktitle}{\emph{International Conference on Machine Learning}}. PMLR, \bibinfo{pages}{3672--3681}.
\newblock


\bibitem[Hardt et~al\mbox{.}(2016)]%
        {hardt2016strategic}
\bibfield{author}{\bibinfo{person}{Moritz Hardt}, \bibinfo{person}{Nimrod Megiddo}, \bibinfo{person}{Christos Papadimitriou}, {and} \bibinfo{person}{Mary Wootters}.} \bibinfo{year}{2016}\natexlab{}.
\newblock \showarticletitle{Strategic classification}. In \bibinfo{booktitle}{\emph{Proceedings of the 2016 ACM conference on Innovations in Theoretical Computer Science (ITCS)}}. \bibinfo{pages}{111--122}.
\newblock


\bibitem[Harris et~al\mbox{.}(2023)]%
        {harris2023strategic}
\bibfield{author}{\bibinfo{person}{Keegan Harris}, \bibinfo{person}{Chara Podimata}, {and} \bibinfo{person}{Steven~Z Wu}.} \bibinfo{year}{2023}\natexlab{}.
\newblock \showarticletitle{Strategic apple tasting}.
\newblock \bibinfo{journal}{\emph{Advances in Neural Information Processing Systems}}  \bibinfo{volume}{36} (\bibinfo{year}{2023}), \bibinfo{pages}{79918--79945}.
\newblock


\bibitem[H{\k{a}}z{\l}a et~al\mbox{.}(2019)]%
        {hazla2019geometric}
\bibfield{author}{\bibinfo{person}{Jan H{\k{a}}z{\l}a}, \bibinfo{person}{Yan Jin}, \bibinfo{person}{Elchanan Mossel}, {and} \bibinfo{person}{Govind Ramnarayan}.} \bibinfo{year}{2019}\natexlab{}.
\newblock \showarticletitle{A geometric model of opinion polarization}.
\newblock \bibinfo{journal}{\emph{arXiv preprint arXiv:1910.05274}} (\bibinfo{year}{2019}).
\newblock


\bibitem[Hegselmann et~al\mbox{.}(2002)]%
        {hegselmann2002opinion}
\bibfield{author}{\bibinfo{person}{Rainer Hegselmann}, \bibinfo{person}{Ulrich Krause}, {et~al\mbox{.}}} \bibinfo{year}{2002}\natexlab{}.
\newblock \showarticletitle{Opinion dynamics and bounded confidence models, analysis, and simulation}.
\newblock \bibinfo{journal}{\emph{Journal of artificial societies and social simulation}} \bibinfo{volume}{5}, \bibinfo{number}{3} (\bibinfo{year}{2002}).
\newblock


\bibitem[Irfan and Ortiz(2014)]%
        {irfan-2014}
\bibfield{author}{\bibinfo{person}{Mohammad~T. Irfan} {and} \bibinfo{person}{Luis~E. Ortiz}.} \bibinfo{year}{2014}\natexlab{}.
\newblock \showarticletitle{On influence, stable behavior, and the most influential individuals in networks: A game-theoretic approach}.
\newblock \bibinfo{journal}{\emph{Artificial Intelligence}}  \bibinfo{volume}{215} (\bibinfo{year}{2014}), \bibinfo{pages}{79--119}.
\newblock
\showISSN{0004-3702}
\href{https://doi.org/10.1016/j.artint.2014.06.004}{doi:\nolinkurl{10.1016/j.artint.2014.06.004}}


\bibitem[Jalan and Chakrabarti(2024)]%
        {jalan-chakrabarti-2024}
\bibfield{author}{\bibinfo{person}{Akhil Jalan} {and} \bibinfo{person}{Deepayan Chakrabarti}.} \bibinfo{year}{2024}\natexlab{}.
\newblock \bibinfo{title}{Strategic Negotiations in Endogenous Network Formation}.
\newblock
\showeprint[arxiv]{2402.08779}~[math.OC]
\urldef\tempurl%
\url{https://arxiv.org/abs/2402.08779}
\showURL{%
\tempurl}


\bibitem[Jalan et~al\mbox{.}(2024)]%
        {jalan-2023}
\bibfield{author}{\bibinfo{person}{Akhil Jalan}, \bibinfo{person}{Deepayan Chakrabarti}, {and} \bibinfo{person}{Purnamrita Sarkar}.} \bibinfo{year}{2024}\natexlab{}.
\newblock \showarticletitle{Incentive-Aware Models of Financial Networks}.
\newblock \bibinfo{journal}{\emph{Operations Research}} \bibinfo{volume}{0}, \bibinfo{number}{0} (\bibinfo{year}{2024}).
\newblock
\href{https://doi.org/10.1287/opre.2022.0678}{doi:\nolinkurl{10.1287/opre.2022.0678}}


\bibitem[Kapoor et~al\mbox{.}(2019)]%
        {kapoor2019corruption}
\bibfield{author}{\bibinfo{person}{Sayash Kapoor}, \bibinfo{person}{Kumar~Kshitij Patel}, {and} \bibinfo{person}{Purushottam Kar}.} \bibinfo{year}{2019}\natexlab{}.
\newblock \showarticletitle{Corruption-tolerant bandit learning}.
\newblock \bibinfo{journal}{\emph{Machine Learning}} \bibinfo{volume}{108}, \bibinfo{number}{4} (\bibinfo{year}{2019}), \bibinfo{pages}{687--715}.
\newblock


\bibitem[Kearns et~al\mbox{.}(2001)]%
        {kearns2001graphical}
\bibfield{author}{\bibinfo{person}{Michael Kearns}, \bibinfo{person}{Michael~L Littman}, {and} \bibinfo{person}{Satinder Singh}.} \bibinfo{year}{2001}\natexlab{}.
\newblock \showarticletitle{Graphical models for game theory}. In \bibinfo{booktitle}{\emph{Proceedings of the Seventeenth conference on Uncertainty in artificial intelligence}}. \bibinfo{pages}{253--260}.
\newblock


\bibitem[Kolumbus et~al\mbox{.}(2023)]%
        {kolumbus2023asynchronous}
\bibfield{author}{\bibinfo{person}{Yoav Kolumbus}, \bibinfo{person}{Menahem Levy}, {and} \bibinfo{person}{Noam Nisan}.} \bibinfo{year}{2023}\natexlab{}.
\newblock \showarticletitle{Asynchronous proportional response dynamics: convergence in markets with adversarial scheduling}.
\newblock \bibinfo{journal}{\emph{Advances in Neural Information Processing Systems}}  \bibinfo{volume}{36} (\bibinfo{year}{2023}), \bibinfo{pages}{25409--25434}.
\newblock


\bibitem[Kolumbus and Nisan(2022)]%
        {kolumbus2022auctions}
\bibfield{author}{\bibinfo{person}{Yoav Kolumbus} {and} \bibinfo{person}{Noam Nisan}.} \bibinfo{year}{2022}\natexlab{}.
\newblock \showarticletitle{Auctions between regret-minimizing agents}. In \bibinfo{booktitle}{\emph{Proceedings of the ACM Web Conference 2022}}. \bibinfo{pages}{100--111}.
\newblock


\bibitem[Leng et~al\mbox{.}(2020)]%
        {leng-2020-learning}
\bibfield{author}{\bibinfo{person}{Yan Leng}, \bibinfo{person}{Xiaowen Dong}, \bibinfo{person}{Junfeng Wu}, {and} \bibinfo{person}{Alex Pentland}.} \bibinfo{year}{2020}\natexlab{}.
\newblock \showarticletitle{Learning quadratic games on networks}. In \bibinfo{booktitle}{\emph{International Conference on Machine Learning}}. \bibinfo{pages}{5820--5830}.
\newblock


\bibitem[Marcus et~al\mbox{.}(2022)]%
        {marcus2022interlacing}
\bibfield{author}{\bibinfo{person}{Adam~W Marcus}, \bibinfo{person}{Daniel~A Spielman}, {and} \bibinfo{person}{Nikhil Srivastava}.} \bibinfo{year}{2022}\natexlab{}.
\newblock \showarticletitle{Interlacing families III: Sharper restricted invertibility estimates}.
\newblock \bibinfo{journal}{\emph{Israel Journal of Mathematics}} (\bibinfo{year}{2022}), \bibinfo{pages}{1--28}.
\newblock


\bibitem[Mueller(2018)]%
        {mueller2018united}
\bibfield{author}{\bibinfo{person}{Robert~S Mueller}.} \bibinfo{year}{2018}\natexlab{}.
\newblock \showarticletitle{United States of America V}.
\newblock \bibinfo{journal}{\emph{Internet Research Agency. Case}} (\bibinfo{year}{2018}).
\newblock


\bibitem[Musco et~al\mbox{.}(2018a)]%
        {musco2018minimizing}
\bibfield{author}{\bibinfo{person}{Cameron Musco}, \bibinfo{person}{Christopher Musco}, {and} \bibinfo{person}{Charalampos~E Tsourakakis}.} \bibinfo{year}{2018}\natexlab{a}.
\newblock \showarticletitle{Minimizing polarization and disagreement in social networks}. In \bibinfo{booktitle}{\emph{Proceedings of the 2018 world wide web conference}}. \bibinfo{pages}{369--378}.
\newblock


\bibitem[Musco et~al\mbox{.}(2018b)]%
        {Musco2018}
\bibfield{author}{\bibinfo{person}{Cameron Musco}, \bibinfo{person}{Christopher Musco}, {and} \bibinfo{person}{Charalampos~E Tsourakakis}.} \bibinfo{year}{2018}\natexlab{b}.
\newblock \showarticletitle{Minimizing polarization and disagreement in social networks}. In \bibinfo{booktitle}{\emph{Proceedings of the 2018 World Wide Web Conference}}. \bibinfo{pages}{369--378}.
\newblock


\bibitem[Nedi{\'c} and Touri(2012)]%
        {nedic2012}
\bibfield{author}{\bibinfo{person}{Angelia Nedi{\'c}} {and} \bibinfo{person}{Behrouz Touri}.} \bibinfo{year}{2012}\natexlab{}.
\newblock \showarticletitle{Multi-dimensional hegselmann-krause dynamics}. In \bibinfo{booktitle}{\emph{2012 IEEE 51st IEEE Conference on Decision and Control (CDC)}}. IEEE, \bibinfo{pages}{68--73}.
\newblock


\bibitem[Nguyen et~al\mbox{.}(2019)]%
        {nguyen2019deception}
\bibfield{author}{\bibinfo{person}{Thanh~H Nguyen}, \bibinfo{person}{Yongzhao Wang}, \bibinfo{person}{Arunesh Sinha}, {and} \bibinfo{person}{Michael~P Wellman}.} \bibinfo{year}{2019}\natexlab{}.
\newblock \showarticletitle{Deception in finitely repeated security games}. In \bibinfo{booktitle}{\emph{Proceedings of the AAAI Conference on Artificial Intelligence}}, Vol.~\bibinfo{volume}{33}. \bibinfo{pages}{2133--2140}.
\newblock


\bibitem[Pariser(2011)]%
        {pariser2011filter}
\bibfield{author}{\bibinfo{person}{Eli Pariser}.} \bibinfo{year}{2011}\natexlab{}.
\newblock \bibinfo{booktitle}{\emph{The filter bubble: How the new personalized web is changing what we read and how we think}}.
\newblock \bibinfo{publisher}{Penguin}.
\newblock


\bibitem[Racz and Rigobon(2022)]%
        {racz2022towards}
\bibfield{author}{\bibinfo{person}{Miklos~Z Racz} {and} \bibinfo{person}{Daniel~E Rigobon}.} \bibinfo{year}{2022}\natexlab{}.
\newblock \showarticletitle{Towards Consensus: Reducing Polarization by Perturbing Social Networks}.
\newblock \bibinfo{journal}{\emph{arXiv preprint arXiv:2206.08996}} (\bibinfo{year}{2022}).
\newblock


\bibitem[R{\'a}cz and Rigobon(2023)]%
        {racz2023towards}
\bibfield{author}{\bibinfo{person}{Miklos~Z R{\'a}cz} {and} \bibinfo{person}{Daniel~E Rigobon}.} \bibinfo{year}{2023}\natexlab{}.
\newblock \showarticletitle{Towards consensus: Reducing polarization by perturbing social networks}.
\newblock \bibinfo{journal}{\emph{IEEE Transactions on Network Science and Engineering}} \bibinfo{volume}{10}, \bibinfo{number}{6} (\bibinfo{year}{2023}), \bibinfo{pages}{3450--3464}.
\newblock


\bibitem[Ristache et~al\mbox{.}(2024)]%
        {tsourakakis-2024}
\bibfield{author}{\bibinfo{person}{Dragos Ristache}, \bibinfo{person}{Fabian Spaeh}, {and} \bibinfo{person}{Charalampos~E Tsourakakis}.} \bibinfo{year}{2024}\natexlab{}.
\newblock \showarticletitle{Wiser than the Wisest of Crowds: The Asch Effect and Polarization Revisited}. In \bibinfo{booktitle}{\emph{Joint European Conference on Machine Learning and Knowledge Discovery in Databases}}. Springer, \bibinfo{pages}{440--458}.
\newblock


\bibitem[Ristache et~al\mbox{.}(2025)]%
        {ristache2025countering}
\bibfield{author}{\bibinfo{person}{Dragos Ristache}, \bibinfo{person}{Fabian Spaeh}, {and} \bibinfo{person}{Charalampos~E Tsourakakis}.} \bibinfo{year}{2025}\natexlab{}.
\newblock \showarticletitle{Countering Election Sway: Strategic Algorithms in Friedkin-Johnsen Dynamics}.
\newblock \bibinfo{journal}{\emph{arXiv preprint arXiv:2502.01874}} (\bibinfo{year}{2025}).
\newblock


\bibitem[Rossi et~al\mbox{.}(2022)]%
        {rossi2022learning}
\bibfield{author}{\bibinfo{person}{Emanuele Rossi}, \bibinfo{person}{Federico Monti}, \bibinfo{person}{Yan Leng}, \bibinfo{person}{Michael Bronstein}, {and} \bibinfo{person}{Xiaowen Dong}.} \bibinfo{year}{2022}\natexlab{}.
\newblock \showarticletitle{Learning to infer structures of network games}. In \bibinfo{booktitle}{\emph{International Conference on Machine Learning}}. PMLR, \bibinfo{pages}{18809--18827}.
\newblock


\bibitem[Roughgarden(2005)]%
        {roughgarden2005selfish}
\bibfield{author}{\bibinfo{person}{Tim Roughgarden}.} \bibinfo{year}{2005}\natexlab{}.
\newblock \bibinfo{booktitle}{\emph{Selfish routing and the price of anarchy}}.
\newblock \bibinfo{publisher}{MIT press}.
\newblock


\bibitem[Roughgarden and Schoppmann(2011)]%
        {roughgarden2011local}
\bibfield{author}{\bibinfo{person}{Tim Roughgarden} {and} \bibinfo{person}{Florian Schoppmann}.} \bibinfo{year}{2011}\natexlab{}.
\newblock \showarticletitle{Local smoothness and the price of anarchy in atomic splittable congestion games}. In \bibinfo{booktitle}{\emph{Proceedings of the Twenty-Second Annual ACM-SIAM Symposium on Discrete Algorithms}}. SIAM, \bibinfo{pages}{255--267}.
\newblock


\bibitem[Russo(2023)]%
        {russo2023analysis}
\bibfield{author}{\bibinfo{person}{Alessio Russo}.} \bibinfo{year}{2023}\natexlab{}.
\newblock \showarticletitle{Analysis and detectability of offline data poisoning attacks on linear dynamical systems}. In \bibinfo{booktitle}{\emph{Learning for Dynamics and Control Conference}}. PMLR, \bibinfo{pages}{1086--1098}.
\newblock


\bibitem[Shearer and Mitchell(2021)]%
        {shearer2021news}
\bibfield{author}{\bibinfo{person}{Elisa Shearer} {and} \bibinfo{person}{Amy Mitchell}.} \bibinfo{year}{2021}\natexlab{}.
\newblock \showarticletitle{News use across social media platforms in 2020}.
\newblock  (\bibinfo{year}{2021}).
\newblock


\bibitem[Tardos(2004)]%
        {tardos-2004}
\bibfield{author}{\bibinfo{person}{Eva Tardos}.} \bibinfo{year}{2004}\natexlab{}.
\newblock \showarticletitle{Network games}. In \bibinfo{booktitle}{\emph{Proceedings of the thirty-sixth annual ACM Symposium on Theory of Computing (STOC)}}. \bibinfo{pages}{341--342}.
\newblock


\bibitem[{Twitter, Inc.}(2019)]%
        {twitter2019hongkong}
\bibfield{author}{\bibinfo{person}{{Twitter, Inc.}}} \bibinfo{year}{2019}\natexlab{}.
\newblock \bibinfo{title}{Information Operations Directed at Hong Kong}.
\newblock
\urldef\tempurl%
\url{https://blog.x.com/en_us/topics/company/2019/information_operations_directed_at_Hong_Kong}
\showURL{%
\tempurl}
\newblock
\shownote{Accessed: 2024-12-27}.


\bibitem[Wang and Kleinberg(2024)]%
        {wang2024relationship}
\bibfield{author}{\bibinfo{person}{Yanbang Wang} {and} \bibinfo{person}{Jon Kleinberg}.} \bibinfo{year}{2024}\natexlab{}.
\newblock \showarticletitle{On the relationship between relevance and conflict in online social link recommendations}.
\newblock \bibinfo{journal}{\emph{Advances in Neural Information Processing Systems}}  \bibinfo{volume}{36} (\bibinfo{year}{2024}).
\newblock


\bibitem[Young(2006)]%
        {young2006diffusion}
\bibfield{author}{\bibinfo{person}{H~Peyton Young}.} \bibinfo{year}{2006}\natexlab{}.
\newblock \showarticletitle{The diffusion of innovations in social networks}.
\newblock \bibinfo{journal}{\emph{The economy as an evolving complex system III: Current perspectives and future directions}}  \bibinfo{volume}{267} (\bibinfo{year}{2006}), \bibinfo{pages}{39}.
\newblock


\bibitem[Zampetakis(2020)]%
        {zampetakis2020statistics}
\bibfield{author}{\bibinfo{person}{Emmanouil Zampetakis}.} \bibinfo{year}{2020}\natexlab{}.
\newblock \emph{\bibinfo{title}{Statistics in high dimensions without IID samples: truncated statistics and minimax optimization}}.
\newblock \bibinfo{thesistype}{Ph.\,D. Dissertation}. \bibinfo{school}{Massachusetts Institute of Technology}.
\newblock


\end{thebibliography}

\appendix

\section{Proofs} \label{app:proofs}

\paragraph{Proof of \cref{theorem:psne}}

Consider agent $i \in S$. To calculate the best-response $\bs_i^\prime$ of $i$ in response to $\bs_{-i}^{\prime}$, we analyze derivatives of its cost function with respect to $\bs^\prime$. Since the equilibrium $\bz^\prime$ is $\bz^\prime = B \bs^\prime$, we have: 
\begin{align*}
c_i(\bz^\prime) &= (1 - \alpha_i) \sum_{j \sim i}
w_{ij} (\bz_i^\prime - \bz_j^\prime)^2 + \alpha_i (\bz_i^\prime - \bs_i)^2 \\
c_i(\bs^\prime) &= (1 - \alpha_i) \sum_{j \sim i}
w_{ij} ((\be_i - \be_j)^T B \bs^\prime)^2 
+ \alpha_i (\be_i^T (B \bs^\prime - \bs))^2 \\
&=  (1 - \alpha_i) \sum_{j \sim i}
w_{ij} (\bs^\prime)^T (B^T (\be_i - \be_j) (\be_i - \be_j)^T B) (\bs^\prime) \\
&+ \alpha_i ((\bs^\prime)^T B^T e_i e_i^T B \bs^\prime 
- 2 (\bs^\prime)^T B^T e_i e_i^T \bs 
+ \bs^T e_i e_i^T \bs) \\
\nabla_{\bs^\prime} c_i(\bs^\prime) 
&= (1 - \alpha_i) \sum_{j \sim i}
w_{ij} 2 (B^T (\be_i - \be_j) (\be_i - \be_j)^T B) (\bs^\prime) 
\\ &  + \alpha_i (2 B^T \be_i \be_i^T B \bs^\prime 
- 2 B^T \be_i \be_i^T \bs), \\
\nabla_{\bs^\prime}^2 c_i(\bs^\prime) &= 2 (1 - \alpha_i) B^T \bigg[\sum_{j \sim i}
w_{ij}  (\be_i - \be_j) (\be_i - \be_j)^T \bigg] B
+2  \alpha_i B^T \be_i \be_i^T B.
 \end{align*}
 Let $L_i \in \RR^{n \times n}$ be: 
 \begin{align*}
L_i := \sum_{j \sim i} w_{ij} (\be_i - \be_j) (\be_i - \be_j)^T  .
\end{align*}
Notice that $L_i$ is precisely the Laplacian of the graph when all edges not incident to $i$ are equal to zero. Therefore $L_i \succeq 0$. Since $\be_i \be_i^T \succeq 0$, the Hessian of $c_i$ with respect to $\bs^\prime$ is PSD. In particular, its $(i,i)$ entry is non-negative, so $\frac{\del^2 c_i(\bs^\prime)}{\del (\bs_i^\prime)^2} \geq 0$, and hence the optimal $\bs_i^\prime$ is at the critical point. This is given as: 
\begin{align*}
0 &= \frac{1}{2} \frac{\del}{\del \bs_i^\prime} c_i(\bs^\prime) \\
&= \be_i^T (1 - \alpha_i) B^T \bigg[\sum_{j \sim i}
w_{ij} (\be_i - \be_j) (\be_i - \be_j)^T \bigg] B \bs^\prime 
\\ & + \be_i^T \alpha_i (B^T \be_i \be_i^T B s^\prime - B^T \be_i \be_i^T \bs) \\
&= (1 - \alpha_i) \be_i^T B^T L_i B \bs^\prime 
+ \be_i^T \alpha_i (B^T \be_i \be_i^T B \bs^\prime - B^T \be_i \be_i^T \bs).
\end{align*}
The above display gives the solution for $\bs_i^\prime$ in terms of all entries of $s^\prime$. Assembling the critical points into a linear system, we obtain precisely that for all $i \in S$, $\be_i^T \TT_i \bs^\prime = \by_i$. Since $\bs_j^\prime = \bs_j$ for $j \not \in S$, the overall linear system describes the Nash equilibria. 

\paragraph{Proof of \cref{prop:blockmodel}}

    Let $V_1$ correspond to the vertex set for community 1 and $V_2$ correspond to the vertex set for community 2. Let $I_{q}$ denote the $q \times q$ identity matrix, and $a_i = \abs{V_i \cap S}$ for $i = 1, 2$.
    Then $$X_S^T X_S = \begin{pmatrix} I_{a_1} & 0 \\ 0 & I_{a_2} \end{pmatrix}.$$ 

    Therefore $\lambda_{\max} (X_S^TX_S)= \max \{ |V_1 \cap S|, |V_2 \cap S| \}$ and $\lambda_{\min} (X_S^T X_S) = \min \{ |V_1 \cap S|, |V_1 \cap S| \}$.

    First, we determine sufficient ranges of $\gamma$ and the value of $\Xi_\gamma$: Let $S$ be such that $|S| = \gamma n$. We have the following options: 

    \begin{itemize}
        \item $S \subseteq V_1$. Then $\lambda_{\max} (X_S^T X_S) = \gamma n$.
        \item $S \subseteq V_2$. Then $\lambda_{\max} (X_S^T X_S) = \gamma n$.
        \item $V_1 \subseteq S$. Then $\lambda_{\max} (X_S^T X_S)$ lies between $\gamma n /2$ and $\gamma n$. 
        \item $V_2 \subseteq S$. Then $\lambda_{\max} (X_S^T X_S)$ lies between $\gamma n /2$ and $\gamma n$. 
        \item If $S$ lies partially in $V_1$ and $V_2$, then $\lambda_{\max} (X_S^T X_S) = \max \{(1 - t)\gamma n, t \gamma n\}$ for some $t \in [0, 1]$. Again, this is upper bounded by $\gamma n$.   
    \end{itemize}

    The above yield $\Xi_\gamma = \gamma n$. To determine $\xi_{1 - \gamma}$ we let $S$ be such that $|S| = (1 - \gamma) n$. We have the following options: 

    \begin{itemize}
        \item $S \subseteq V_1$. Then $\lambda_{\min} (X_S^T X_S) = 0$. 
        \item $S \subseteq V_2$. Then $\lambda_{\min} (X_S^T X_S) = 0$. 
        \item $V_1 \subseteq S$. Then $\lambda_{\min} (X_S^T X_S) = \min \{ n_1, (1 - \gamma) n - n_1 \} = n_1 \ge (1 - \gamma)n - n_2$ since always $n_1 \ge (1 - \gamma) n/2$. 
        \item $V_2 \subseteq S$. Then $\lambda_{\min} (X_S^T X_S) = \min \{ n_2, (1 - \gamma)n  - n_2 \} = (1 - \gamma)n - n_2$ since $n_2 \le (1 - \gamma)n/2$. 
        \item If $S$ lies partially in $V_1$ and $V_2$, then $\lambda_{\min} (X_S^T X_S) = \min \{(1 - t)(1 - \gamma) n, t (1 - \gamma) n\}$ for some $t \in [0, 1]$. Again, this is lower bounded by $( 1-  \gamma) n - n_2$.
    \end{itemize}

    Therefore, for either $1 - \gamma \le n_1 /n$ or $1 - \gamma \le n_2 / n$ we have that $\xi_{1 - \gamma} = (1 - \gamma) n - n_2$. The final inequality corresponds to 

    \begin{equation*}
        4 \sqrt {\frac {\Xi_\gamma} {\xi_{1 - \gamma}}} < 1 \iff \gamma < \frac {1} {17} - \frac {n_2} {17n}
    \end{equation*}

    Combining the above we get two systems of inequalities. The first one corresponds to $1 - \frac {n_1} {n} \le \gamma < \frac {1} {17} - \frac {n_2} {17n}$ which holds for $n_2 < 1/18$ which is impossible since $n_2 \ge 1$. The second one corresponds to $1 - \frac {n_2} n \le \gamma < \frac {1} {17} - \frac {n_2} {17n}$ which holds for $n_2 > 16/18$, which is always true.



\paragraph{Proof of Theorem~\ref{thrm:pom_shared_alpha}}
First, we set $\tilde \alpha = \alpha / (1 - \alpha)$. By substituting $\bz = B \bs$ we can show by straightforward algebra that $C(\bz) / (1 - \alpha) = \bs^T Q \bs$ where $Q \succeq 0$ with 

\begin{align}
    Q = B L B + \tilde \alpha (I - 2B + B^2)
\end{align}
Since $Q \succeq 0$, it has eigendecomposition $Q = U \Lambda_Q U^T$. Moreover, $U$ is precisely the matrix of eigenvectors for the Laplacian. The eigenvalues of $Q$ can be shown to be $\tilde \alpha^2 / (\lambda_i + \tilde \alpha)$. Therefore, $C(z) = \bs^T Q \bs \ge \frac {\tilde \alpha^2} {\lambda_n + \tilde \alpha} \norm \bs \norm_2^2$. 

Next, let $\diag (B)$ be the diagonal matrix with entries $B_{ii}$ and $\widetilde {\diag (B)}$ be the diagonal matrix with elements $B_{ii}$. It is easy to show that $\bs^\prime = (1/\tilde \alpha) B^{-1} \widetilde {\diag (B)} \bs$ and $\bz^\prime = (1 /\tilde \alpha) \widetilde{\diag (B)} \bs$, which similarly implies (after algebraic operations) that $C(\bz^\prime) / (1 - \alpha) = \bs^T Q^\prime \bs$ where: 

\begin{align*}
    Q^\prime & := \frac {1} {\tilde \alpha^2} \widetilde{\diag (B)} L \widetilde {\diag (B)} + \frac {1} {\tilde \alpha} B^{-1} \left ( \widetilde{\diag (B)} \right )^2 B^{-1} \\ & - 2 \frac {1} {\tilde \alpha} B^{-1} \left ( \widetilde{\diag (B)} \right )^2 + \frac {1} {\tilde \alpha} \left ( \widetilde{\diag (B)} \right )^2
\end{align*}

Note that $Q^\prime$ cannot be diagonalized since, in general, $\widetilde{\diag (B)}$ has a different eigenbasis than $L$. However, we note that: 
\begin{align} \label{eq:ineq_Bi}
    \norm \widetilde {\diag (B)} \norm_2 & = \max_i B_{ii} \le \norm B \norm_2 = 1, \\
    \norm B^{-1} \norm_2 & = \max_i \frac {\lambda_i + \tilde \alpha} {\tilde \alpha} = \frac {\lambda_n + \tilde \alpha} {\tilde \alpha}. \label{eq:ineq_Binv}
\end{align}
By the triangle inequality, the Cauchy-Schwarz inequality, and \cref{eq:ineq_Bi,eq:ineq_Binv}, we have that: 
\begin{align*}
    \norm Q^\prime \norm_2 & \le \frac {1} {\tilde \alpha^2} \norm L \norm_2 \left ( \norm \widetilde {\diag (B)} \norm_2 \right )^2 + \frac {1} {\tilde \alpha} \left ( \norm \widetilde {\diag (B)} \norm_2 \right )^2 \norm B^{-1} \norm_2^2 \\ & +  \frac 2 {\tilde \alpha} \norm B^{-1} \norm_2  +  \frac {1} {\tilde \alpha} \left ( \norm \widetilde {\diag (B)} \norm_2 \right )^2 
    \\ & \le \frac {(\lambda_n + 4 \tilde \alpha) (\lambda_n + \tilde \alpha)} {\tilde \alpha^3}.
\end{align*}

Therefore $C(\bz^\prime) / (1 - \alpha) \le \frac {(\lambda_n + 4 \tilde \alpha) (\lambda_n + \tilde \alpha)} {\tilde \alpha^3} \norm \bs \norm_2^2$. Hence,
\begin{align}
    \frac {C(\bz^\prime)} {C(\bz)} \le \frac {(\lambda_n + 4 \tilde \alpha) (\lambda_n + \tilde \alpha)^2} {\tilde \alpha^5}.
\end{align}

Finally, we can simplify: 
\begin{align}
    \frac {(\lambda_n + 4 \tilde \alpha) (\lambda_n + \tilde \alpha)^2} {\tilde \alpha^5} \le \frac {64 (\lambda_n + \tilde \alpha)^3} {\tilde \alpha^5} \le \frac {128 (\max \{ \lambda_n, \tilde \alpha \})^3} {\tilde \alpha^5}.
\end{align}    

\paragraph{Extension to heterogeneous $\alpha_i$}

From Theorem~\ref{thrm:pom_shared_alpha}, we can show that the upper bound is minimized when $\lambda_n = \Theta (\tilde \alpha^3)$ and has a value of $O(1/\tilde \alpha^2)$. As we noted, Theorem~\ref{thrm:pom_shared_alpha} can be written with $d_{\textup{max}}$ in the place of $\lambda_n$ as well. 

Next, we give an easy generalization to the case of differing susceptibility. 
\begin{cor}[Price of Misreporting for Heterogeneous Susceptibility]
If the $\alpha_i$ are differing, let $\alpha_{\min} = \min_i \alpha_i$ and $\alpha_{\max} = \max_j \alpha_j$. Define $\tilde \alpha_{\min} = \frac{\alpha_{\min}}{1 - \alpha_{\max}}, \tilde \alpha_{\max} = \frac{\alpha_{\max}}{1 - \alpha_{\min}}$. The Price of Misreporting is bounded as:
\begin{align*}
        \pom \le \frac {1 -\alpha_{\min}} {1 - \alpha_{\max}}\frac {(\lambda_n + 4 \tilde \alpha_{\max}) (\lambda_n + \tilde \alpha_{\max})^2} {\tilde \alpha_{\min}}. 
    \end{align*}
\end{cor}

Note that $C(\bz^\prime) \le (1 - \alpha_{\min}) (\bz^\prime)^T L \bz^\prime + \alpha_{\max} \norm \bz^\prime - \bs \norm_2^2 = \overline C(\bz^\prime)$, and $C(\bz) \ge (1 - \alpha_{\max}) \bz^T L \bz + \alpha_{\min} \norm \bz - \bs \norm_2^2 = \underline C(\bz)$ where $\alpha_{\min} = \min_{i \in [n]} \alpha_i$, and $\alpha_{\max} = \max_{i \in [n]} \alpha_i$. Then, the same analysis of \cref{thrm:pom_shared_alpha} can be applied, since $\overline C(\bz^\prime) / \underline C(\bz) \ge C(\bz^\prime) / C(\bz)$.

\paragraph{Proof of \cref{prop:b-recovery}}
Let $\widehat{\bm{s}^\prime}$ be as in Algorithm~\ref{alg:learning}, and $\bm{y} = \widehat{\bm{s}^\prime}$. Notice $\bm{y} = \bm{s} + (\bm{s}^\prime - \bs) + (\widehat{\bm{s}^\prime} - \bs^\prime)$. Let $\bw := (\bm{s}^\prime - \bs)$ be the corruption vector due to strategic negotiations and $\bm{r} =  (\widehat{\bm{s}^\prime} - \bs^\prime)$ be the residual vector due to least-squares regression. We claim that $\bm{r} = \bm{0}$, because $A^{-1}$ is full rank and $((I - A) L + A)$ is full rank, so $\widehat{\bm{s}^\prime} = A^{-1} ((I - A) L + A) \bz^\prime = \bs^\prime$. 

Next, we apply Theorem ~\ref{thrm:torrent}. Notice that $\norm \bw \norm_0 \leq Cn$ by assumption. Moreover, $X$ satisfies the SSC and SSS coniditons. Therefore, after $T$ iterations, Algorithm~\ref{alg:learning} obtains $\hat \bs$ such that: 
\[
\norm \hat \bv - \bv \norm_2 \leq \frac{\exp(-cT)}{\sqrt{n}} \norm \bs^\prime - \bs \norm_2.
\]
Therefore, letting $T = C^\prime (\log n)^2$ for large enough constant $C^\prime > 0$, we see that $\norm \hat \bv - \bv \norm_2 \leq O(n^{-\omega(1)})$. Hence $\norm \hat \bs - \bs \norm_2 = \norm X \hat \bv - X \bv \norm_2 \leq \norm X \norm_2 \cdot n^{-\omega(1)}$. Now, let $\bm{u} = \hat \bs - \bs^\prime$. If $i \in [n] \setminus S$, then $\abs{\bm{u}_i} \leq \norm X \norm_2 \cdot n^{-\omega(1)}$. On the other hand for $j \in S$, $\abs{\bm{u}_{j}} \geq \abs{\bm{s}_j - \bm{s}_j^\prime} - \norm X \norm_2 n^{-\omega(1)}$. Therefore the top-$|S|$ entries of $\bm{u}$ recover $S$. 

\paragraph{Proof of \cref{prop:blockmodel_k}}

First, we note that if $V_1, \dots, V_K$ are the vertex sets and $S$ is a set of size $\gamma n$ the maximum eigenvalue equals to $\lambda_{\max} (X_S^T X_S) = \max_{i \in [K]} |V_i \cap S|$ and is always at most $\gamma n$. So $\Xi_\gamma = \gamma n$. If $S$ is a set of size $(1 - \gamma) n$, then the minimum eigenvalue $\lambda_{\min}(X_S^T X_S) = \min_{i \in [K} |V_i \cap S|$ is maximized when $|V_1 \cap S| = \dots = |V_K \cap S| = (1 - \gamma) n / K$ which holds as long as $(1 - \gamma)n / K \le n_K$, so $\xi_{1 - \gamma} = (1 - \gamma)n / K$ as long as $\gamma \ge 1 -  n_K/n  K$. Also the other condition is

    \begin{equation*}
        4 \sqrt {\frac {\Xi_\gamma} {\xi_{1 - \gamma}}} < 1 \iff \gamma < \frac {1} {16K + 1}
    \end{equation*}

    Finally, we must have $1 / (16K + 1) > 1 -  K n_K / K$ which yields $n_K > n/K (16K / (16K + 1))$.

\section{Additional Figures} \label{app:additional_figures}

\begin{figure*}[!h]
    \centering
    \includegraphics[width=0.6\linewidth]{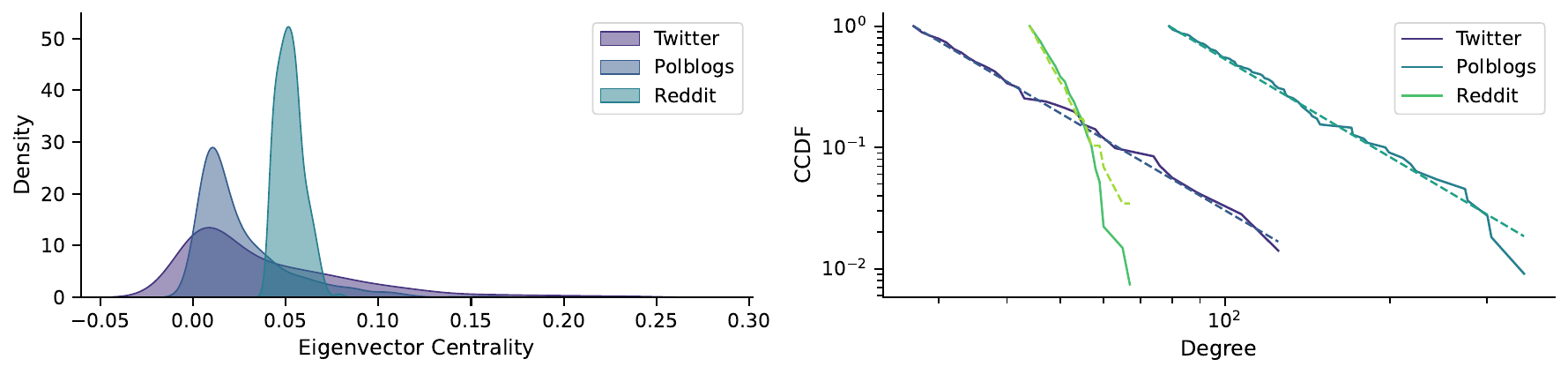}
    \caption{Distribution of centralities and degrees for the datasets}
    \label{fig:network_stats}
    \includegraphics[width=0.6\linewidth]{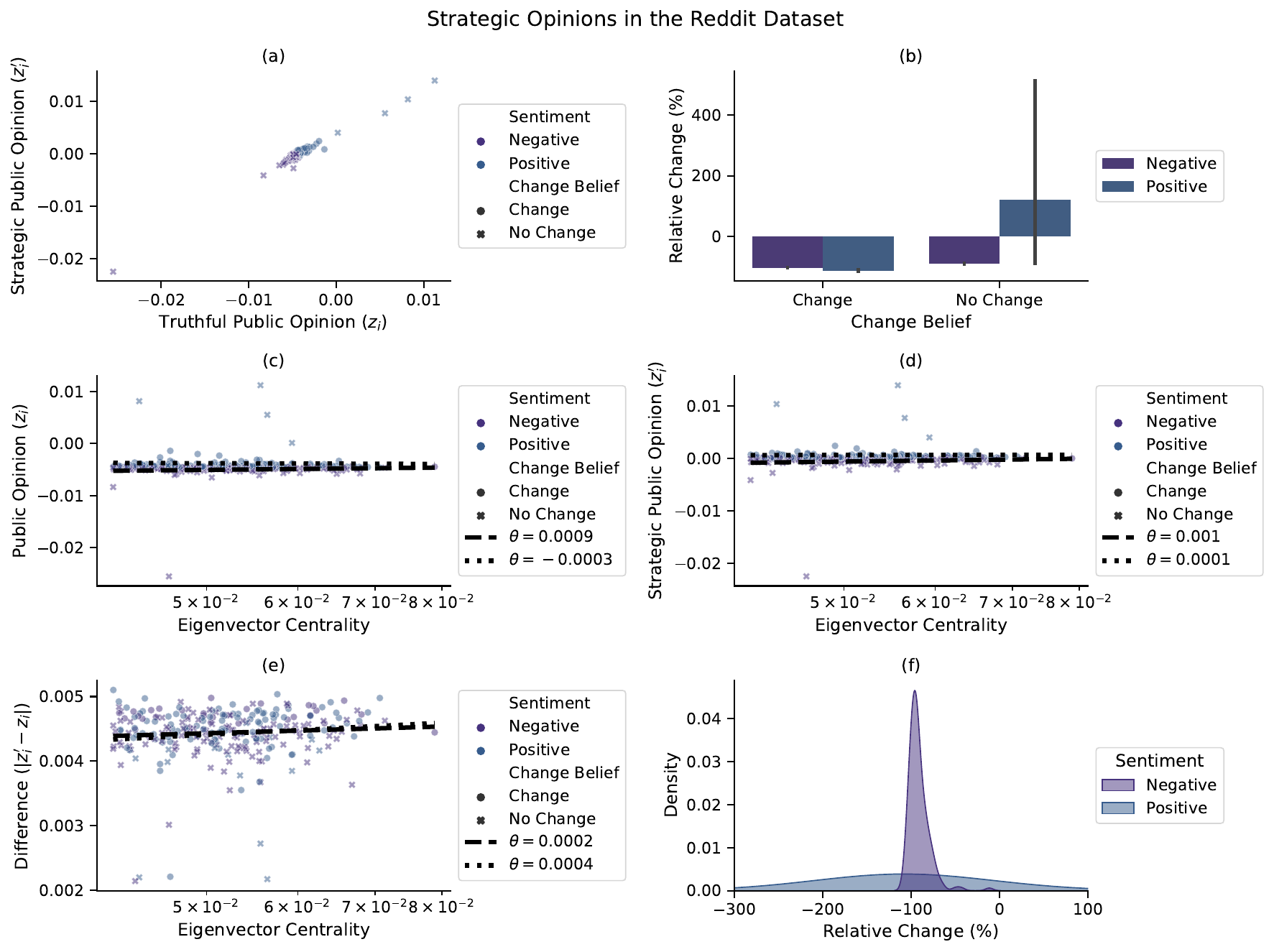}
    \caption{Running the experiments of Figure~\ref{fig:polblogs} for the Reddit dataset.}
    \label{fig:reddit}
    \includegraphics[width=0.6\linewidth]{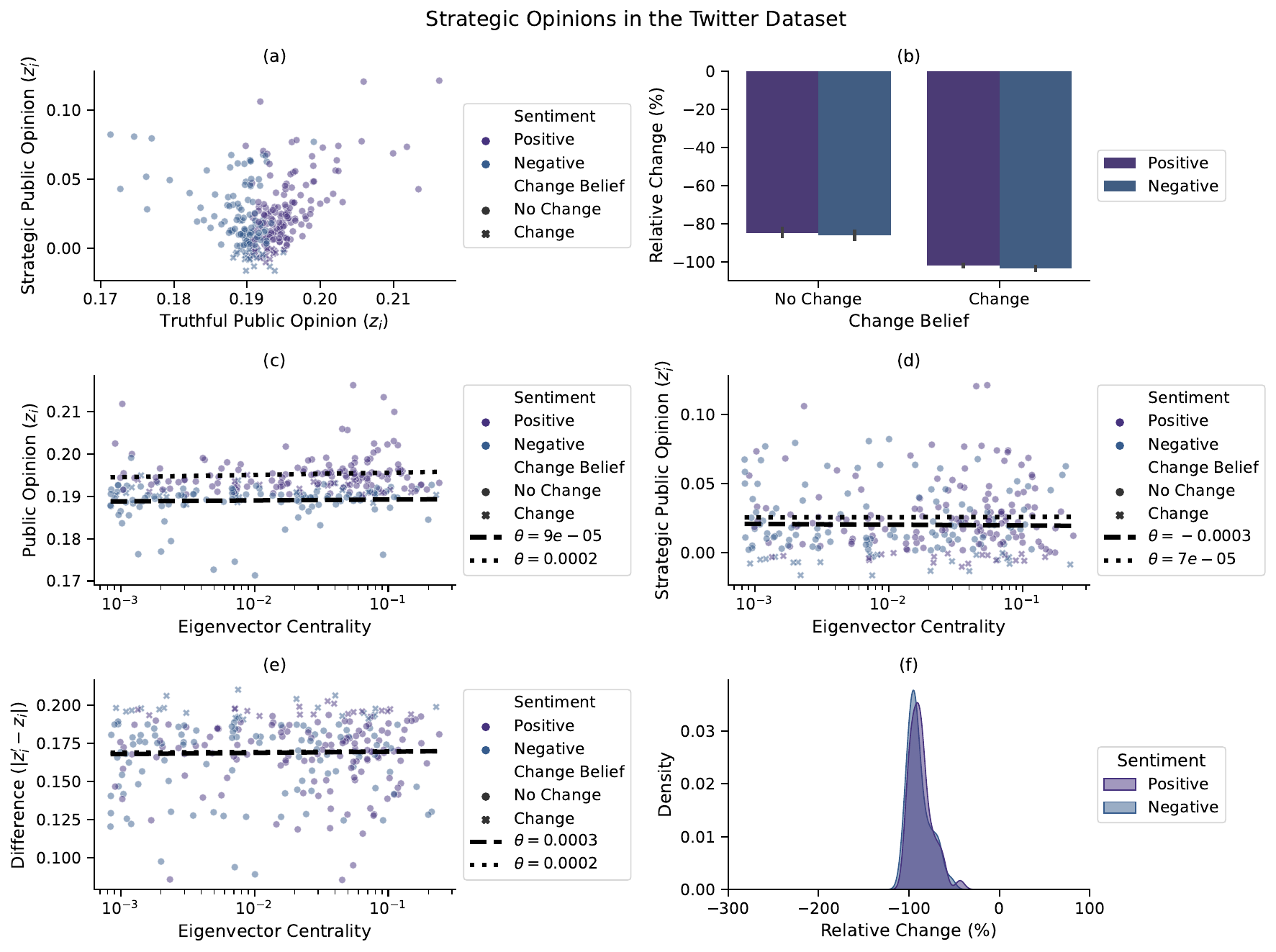}
    \caption{Running the experiments of Figure~\ref{fig:polblogs} for the Twitter dataset.}
    \label{fig:twitter}
\end{figure*}

\end{document}